\documentclass[a4paper,11pt]{article}
\usepackage{jcappub} 
\usepackage{orcidlink}
\usepackage{makecell}
\usepackage{amsmath}
\usepackage{longtable}
\usepackage{multirow}
\usepackage{romannum}
\usepackage{booktabs}  
\usepackage{threeparttable}  
\usepackage{adjustbox}

\usepackage{tikz}
\usepackage{placeins}
\usetikzlibrary{arrows.meta,positioning,calc}

\begin{document}

\title{Wedge-avoidance Fisher Forecasts for Primordial Features 
from Dark-Ages 21-cm Power Spectrum and Bispectrum}

\author[a,b]{Chen Zhang,}
\author[a,b]{Furen Deng,}
\author[a,1]{Yidong Xu,}\footnote{Corresponding author}
\author[a]{Bin Yue,}
\author[a]{Yuewei Wen}
\author[a,b,1]{and Xuelei Chen}

\affiliation[a]{State Key Laboratory of Radio Astronomy and Technology, 
National Astronomical Observatories, CAS,
20A Datun Road, Chaoyang District, Beijing 100101, 
People's Republic of China}

\affiliation[b]{School of Astronomy and Space Science, 
University of Chinese Academy of Sciences,
Beijing 100049, People's Republic of China}

\emailAdd{xuyd@nao.cas.cn}
\emailAdd{xuelei@cosmology.bao.ac.cn}

\abstract{The Dark Ages offer a uniquely clean window on primordial physics, making the redshifted 21-cm signal a promising probe of inflation models. 
Forecasts for interferometric 21-cm measurements must account for foreground wedge avoidance, which removes a substantial fraction of Fourier modes, and this effect is especially severe at high redshifts. We develop a wedge-aware Fisher framework in which this mode loss is propagated directly into the variances of the anisotropic 21-cm power spectrum and reduced bispectrum. 
As a case study, we apply the method to forecast  
constraints on two inflation models with different oscillatory features, a resonant model and a step model, for a lunar far-side array including thermal noise.
Since these features affect both the power spectrum and the bispectrum, we apply wedge avoidance to both forecasts and compare their constraining power on oscillatory  amplitudes in these   
feature models. We find that wedge avoidance reduces available Fourier modes significantly in both observables, leading to much weaker constraints on inflationary features. 
This framework is broadly applicable to 21-cm power spectrum and bispectrum forecasts across redshifts and is particularly useful in regimes where foreground wedge effect is severe, such as the epoch of reionization and the higher-redshift Dark Ages. 
}

\maketitle
\flushbottom

\section{Introduction}
Current observations of the cosmic microwave background (CMB) anisotropies and the large-scale structures are consistent with the standard cosmological model with a nearly Gaussian and scale-invariant primordial power spectrum \citep{Planck2019,Mergulhao2023,Martinez-Somonte2026}.
However, a number of physically-motivated inflation models predict scale-dependent oscillatory features in the primordial power spectrum, but still not well constrained \citep{Chluba2015}.
Previous \textit{Planck} analyses reported preferences for primordial features \citep{Benetti2013, Hazra2016}. However, a recent reanalysis of \textit{Planck} using updated maps and unbinned spectra found that these preferences were not supported by Bayesian analysis and the global significance was reduced to at most 2.6$\sigma$ \citep{Raffaelli2026}. Although current constraints are 
tight for the searched parameter space \citep{Planck2018,Raffaelli2026}, further improvement from the CMB is ultimately limited by cosmic variance, whereas large-scale structure surveys can continue to improve their sensitivity by accessing a three-dimensional cosmic volume and hence potentially a much larger number of independent modes  \citep{Alvarez2014,Chaussidon2024,Euclid2025}.

An attractive possibility is to search for primordial features 
during the Dark Ages with the redshifted 21-cm hyperfine transition line of neutral hydrogen (HI), which can inherently access to a large number of three-dimensional modes while retaining 
a comparatively clean primordial signal. The Dark Ages, the period between cosmic recombination and the formation of the first luminous sources, are difficult to probe with conventional observations  \citep{Pritchard2011}, but they provide a uniquely clean window for primordial physics  \citep{Lewis2007,Floss:2022grj,Mondal2023,Vanetti:2025aca,Smith:2025udg}. 
At this stage, the matter and radiation fields still preserve information about the primordial conditions and the initial dark-matter distribution, and the redshifted 21-cm line 
offers a unique access to this epoch. Complementary to the CMB, which provides a two-dimensional snapshot on the last-scattering surface, the redshifted 21-cm line enables tomographic, three-dimensional mapping of HI across cosmic time, dramatically increasing the number of accessible Fourier modes and hence the achievable precision  \citep{Loeb2003,Mao2008,Silk2020}.

Motivated by this potential, a number of works have explored the impact and observability of primordial non-Gaussianity (PNG) in 21-cm signals, either in the power spectrum or in the bispectrum, in order to constrain the primordial physics. Early studies suggested that future low-frequency radio arrays could detect secondary non-Gaussian signals generated by gravitational collapse \citep{Pillepich2006}, and under ideal cosmic-variance-limited conditions, one could detect the PNG 
at the level of \(f_{\rm NL} \sim \mathcal{O}(1)\) using the 21-cm signal \citep{Pillepich2006}. More recent forecasts indicate that an HI galaxy redshift survey with the full-scale Square Kilometre Array (SKA) can reach \(\sigma\!\left(f_{\rm NL}^{\rm loc}\right) \simeq 1.5\) \citep{Camera2014}, while Stage-II 21-cm intensity mapping concepts such as Packed Ultra-wideband Mapping Array (PUMA) may achieve \(\sigma\!\left(f_{\rm NL}^{\rm loc}\right) < 1\) combining 21-cm power spectrum and bispectrum, provided that foregrounds and systematics can be well controlled \citep{Karagiannis2019}. 
In the present work, we focus on a special class of inflation models with oscillatory features, i.e. the resonant model \citep{Chen2008,Flauger2010} and the step model \citep{Adams2001,Chen2006,Chen2008}, for which signals would be suppressed in the CMB due to the projection effect while the three-dimensional (3D) 21-cm observations could potentially be more promising \citep{Xu2016}.

Observing the Dark-Ages 21-cm signal is, however, exceptionally challenging. Ground access to these frequencies is severely limited by the Earth's ionosphere and by strong radio-frequency interferences (RFI). These motivates the idea of building radio interferometer arrays on the lunar far side, a naturally radio-quiet environment. Several concepts, e.g. LARAF \citep{Chen2024},
FARSIDE \citep{Burns2021}, and FarView \citep{Polidan2024,Burns:2026bul}, envision large low-frequency arrays expressly designed to detect the Dark-Ages 21-cm signal with phased construction plan. Even in such a favorable environment, foreground contamination remains a central obstacle and great challenge. In interferometric observations, smooth-spectrum foregrounds will leak into a characteristic region of \((k_\perp,k_\parallel)\) space known as the foreground wedge \citep{Morales:2012kf,Parsons:2012qh,Liu:2014yxa}, in which the Fourier modes are contaminated and may be difficult to be cleaned for cosmological analysis. At the very high redshifts where the observing frequency is lower, the wedge removes a larger fraction of modes than in lower-redshift 21-cm surveys, leading to a correspondingly larger loss of cosmological information \citep{Pober_2025}.
Previous studies predicted promising PNG constraints using 21-cm tomographic observations, while relying on relatively simplified foreground treatments \citep{Pillepich2006,Camera2014,Karagiannis2019}, which suggests the need for more careful and realistic modeling.
Ref. \cite{Karagiannis2019} accounted for foreground filtering and wedge avoidance by restricting the accessible Fourier modes and bispectrum triangle orientations in the forecasts, but their formalism does not propagate the wedge-induced mode loss into the covariance of the bispectrum.

In this work, we develop a wedge-aware Fisher-forecasting framework to constrain primordial oscillatory features 
during the Dark Ages using the anisotropic 
21-cm power spectrum and the anisotropic 
reduced bispectrum, respectively. Specifically, we consider two typical feature templates for inflationary 
models, the \emph{resonant} type and the \emph{step} type,  
and forecast the uncertainties on their oscillatory amplitude parameters, \(f_{\rm res}\) and \(\epsilon_{\rm step}\), respectively, for a virtual lunar far-side array including thermal noise. We also evaluate the total signal-to-noise ratio of each statistic, comparing cases with and without foreground wedge avoidance. The foreground wedge is implemented as a hard mask in the \((k_\perp,k_\parallel)\) space, and the resulting loss of modes is propagated into the variance of each statistic in the Fisher information calculation. 

This paper is organized as follows. Section~\ref{sec:methods} presents the Dark-Ages 21-cm signal model for the power spectrum $P_{21}$ and reduced bispectrum $Q_{21}$, together with the \emph{resonant} and \emph{step} feature models for the primordial term. 
Section~\ref{sec:fisher_formalism} present the standard Fisher forecast formalism for each statistic, and section~\ref{sec:fisher_wedge} gives the wedge-avoidance Fisher formalism,  
where foreground-mode removal and instrumental noise are propagated into the variances. Section~\ref{sec:intrumentations} describes the notional lunar far-side array and the associated thermal noise. Section~\ref{sec:results} reports the main results. We conclude in section~\ref{sec:conclusions} and discuss implications for future observations  
and related high-redshift applications.

\section{Methods} \label{sec:methods}
\subsection{21-cm Power Spectrum and Bispectrum from the Dark Ages}

During the cosmic Dark Ages ($30 \lesssim z \lesssim 1100$), the redshifted 21-cm signals can be measured as the differential brightness temperature with respect to the intrinsic CMB temperature, caused by spin-flip absorption by neutral hydrogen (HI).
In the optically thin limit, which is appropriate for the epoch, it is given by (e.g. \cite{Furlanetto2006})
\begin{equation}
T_{21}(\mathbf x ,z) = \tau(\mathbf x ,z)\, \frac{T_{\rm S}(\mathbf x, z)-T_{\rm CMB}(z)}{1+z},
\end{equation} 
where $T_{\rm CMB}$ is the CMB temperature, $T_{\rm S}$ is the spin temperature of HI, and $\tau$ is the 21-cm optical depth
depending on HI density, spin temperature, and line-of-sight velocity gradient.
Since hydrogen is mostly neutral during the Dark Ages, we write
\begin{equation}
    n_{\rm H}(\mathbf x, z)=\bar n_{\rm H}(z)\left[1+\delta_b(\mathbf x,z)\right],
\end{equation}
and the 21-cm brightness fluctuations are mainly sourced by fluctuations in the baryon density contrast, the gas temperature, and the velocity gradient field.

Following \cite{Ali-Haimoud:2013hpa} and \cite{Munoz2015}, we define $\delta_v(\boldsymbol x) \equiv -(1+z) (\partial v_{\parallel}/\partial r_{\parallel})/H(z)$
where the $\partial v_{\parallel}/\partial r_{\parallel}$ is the gradient of the peculiar velocity along the line of sight
and decompose the baryon density contrast 
into a linearly-evolving term $\delta_b^{(1)}$ and a quadratic term $\delta_b^{(2)}$ which describes the leading non-linear correction arising from gravitational evolution
\begin{equation}
\delta_{\rm b}(\boldsymbol x ,z)=\delta_{\rm b}^{(1)}(\boldsymbol x,z)+\delta_{\rm b}^{(2)}(\boldsymbol x,z)+\mathcal O(\delta_b^{(3)}).
\end{equation}
Considering only scales larger than the baryonic Jeans scale, we assume that the baryonic overdensity follows the evolution of  dark matter overdensity, so that $\delta_b^{(1)}$ grows as the scale factor $a$ and $\delta_b^{(2)}\propto a^2$. Then, to the second order,
the dependence of 21-cm brightness-temperature fluctuations on the baryonic fluctuations can be approximately described by the following effective coefficients \citep{Munoz2015},
\begin{equation}
\alpha \equiv \frac{\partial T_{21}}{\partial \delta_b^{(1)}}, 
\qquad 
\beta \equiv \frac{1}{2}\,\frac{\partial^{2} T_{21}}{\partial (\delta_b^{(1)})^{2}}, 
\qquad 
\gamma \equiv \frac{\partial T_{21}}{\partial \delta_b^{(2)}}.
\end{equation}
Further, we consider a small patch of the sky, across which we can take the line-of-sight direction, $\boldsymbol{\hat n}$, to be constant. Then to the lowest order, the 21-cm brightness-temperature power spectrum can be approximately written as \citep{Munoz2015}
\begin{equation}
P_{21}(\boldsymbol{k},z)
  = \bigl[\alpha + \bar{T}_{21}\,\mu^{2}\bigr]^{2}\,P_{\delta_b}(k,z),
\label{eq:P_21}
\end{equation}
where $P_{\delta_b}$ is the baryon power spectrum, and $\mu = (\boldsymbol{k}\cdot \boldsymbol{\hat n})/|\boldsymbol{k}|$, which relates the peculiar velocity term to the density contrast at linear order, $\delta_v^{(1)}(\boldsymbol{k})=\mu^2\delta_b^{(1)}(\boldsymbol{k})$. On linear scales and during the Dark Ages, baryons trace the cold dark matter, so we set $\delta_b = \delta_{\mathrm{m}} = \delta$
and use the linear matter power spectrum as the baryon power spectrum, setting $P_{\delta_b}(k,z) = P_{\delta}(k,z) = P_{\mathrm{m}}^{\mathrm{lin}}(k,z)$. 

Finally, for the primordial feature models that we study in this paper, the 21-cm brightness-temperature power spectrum can be computed by
\begin{equation}
    P_{21}(\boldsymbol{k},z)
  = \bigl[\alpha + \bar{T}_{21}\,\mu^{2}\bigr]^{2}\,P^{\rm lin}_{\rm m}(k,z)\times 
\frac{\mathcal{P}^{\mathrm{feature}}_{\Phi}(k)}{
\mathcal{P}^{(0)}_{\Phi}(k)},
\end{equation}
where $\mathcal{P}^{\mathrm{feature}}_{\Phi}(k)$ is the dimensionless Bardeen-potential power spectrum of a specific feature  model, and $\mathcal{P}^{(0)}_{\Phi}$ is the standard nearly scale-invariant dimensionless Bardeen-potential power spectrum. During the matter-dominated era, the relation between Bardeen-potential $\Phi$ and curvature perturbation $\mathcal R$ is $\Phi=3\mathcal R /5$, with $\mathcal P_{\mathcal R}^{(0)}(k)=\mathcal A_s (k/k_{\rm p})^{n_s-1}$ and $\mathcal P_{\Phi}^{(0)}(k)=(9/25)\mathcal P_{\mathcal R}^{(0)}(k)$, where $n_s$ is the scalar spectra index, $\mathcal A_s\simeq 2.3\times 10^{-9}$  is the amplitude of scalar perturbation power spectrum, and \(k_{\mathrm p}=0.02\,\mathrm{Mpc}^{-1}\) is a fixed pivot scale. 

In the following, we will work with the anisotropic 21-cm power spectrum and bispectrum in order to properly account for the effect of foreground avoidance on the expected measurement precision.
However, for the purpose of illustrating the oscillatory features on the 21-cm power spectrum and bispectrum in Figures~\ref{fig:resonant_P} -- \ref{fig:step_Q_3}, here we also give spherically averaged expressions. The spherically averaged 21-cm brightness-temperature power spectrum is
\begin{equation}
\begin{aligned}
    P_{21}(k,z)&=\frac{1}{4\pi}\int_0^{2\pi}\mathrm d \phi \int_{-1}^{1}\mathrm d \mu \ P_{21}(\boldsymbol k,z)
    \\
    &=\left(\alpha^2 +\frac{2}{3}\alpha\, \bar T_{21}+\frac{1}{5}\bar T^2_{21}\right) P^{\rm lin}_{\rm m}(k,z)\times 
\frac{\mathcal{P}^{\mathrm{feature}}_{\Phi}(k)}{
\mathcal{P}^{(0)}_{\Phi}(k)}.
\end{aligned}
\end{equation}

The 21-cm brightness-temperature bispectrum is contributed by a primordial non-Gaussianity term, $B_{21}^{\rm I}$, secondary non-gaussianities induced by non-linear gravitational collapse, $B_{21}^{\rm G}$, and  by the nonlinear mapping between the matter fields and the 21-cm brightness-temperature fluctuation, $B_{21}^{\rm nl}$ \citep{Munoz2015},
so that
\begin{equation}
B_{21}(\boldsymbol{k_1},\boldsymbol{k_2},\boldsymbol{k_3};z)=
B_{21}^{\rm I}(\boldsymbol{k_1},\boldsymbol{k_2},\boldsymbol{k_3};z)+
B_{21}^{\rm G}(\boldsymbol{k_1},\boldsymbol{k_2},\boldsymbol{k_3};z)+
B_{21}^{\rm nl}(\boldsymbol{k_1},\boldsymbol{k_2},\boldsymbol{k_3};z).
\label{eq:B_tot}
\end{equation}

\emph{(i)~Primordial non-Gaussianity term $B_{21}^{\rm I}$.}  
For a given type of PNG specified by the initial potential  bispectrum $B_{\Phi}$, the primordial contribution to the 21-cm bispectrum is \citep{Munoz2015},
\begin{equation}
B_{21}^{\rm I}=\bigl[\alpha+\bar{T}_{21}\mu_1^{2}\bigr]
\bigl[\alpha+\bar{T}_{21}\mu_2^{2}\bigr]
\bigl[\alpha+\bar{T}_{21}\mu_3^{2}\bigr]\,
M_1\,M_2\,M_3\,B_\Phi(k_1,k_2,k_3),
\label{B_I}
\end{equation}
where $\mu_i = (\boldsymbol{k_i}\cdot \boldsymbol{\hat n})/k_i$, and
$M_i\!=\!M(k_i,z)\equiv \delta(\boldsymbol k_i,z)/\Phi(k_i)$ relates the density fluctuations to the primordial curvature perturbation in Fourier space, via the Poisson equation. 
The full expression of $M(k_i,z)$ is (e.g.  \cite{Matarrese2008,Karagiannis2018,Dodelson2020}) 
\begin{equation}
    M(k,z)=\frac{2 k^2 c^2}{3\Omega_m H_0^2}T(k)D_{+}(z),
    \label{Mkz}
\end{equation}
where $c$ is the speed of light, 
$T(k)$ is the matter transfer function normalized to unity on large scales, $D_{+}(z)$ is the growth factor of the growing mode of density perturbations, and $D_{+}(z)/a=1$ during the matter domination era. $\Omega_m$ is the present-day matter density parameter and $H_0$ is the Hubble constant.

\emph{(ii)~Non-linear Gravitational growth term $B_{21}^{\rm G}$.}
Even for perfect gaussian initial condition, the non-linear growth of fluctuations induced by gravity will generate non-zero bispectrum, which can be computed from second-order perturbation theory. We define the normalized velocity divergence $\theta\equiv-\nabla_{\rm phys}\cdot\boldsymbol v /H$ during matter-dominated era, so that $\delta_v(\boldsymbol k)=\mu^2 \theta(\boldsymbol k)$. Assuming the linear baryon overdensity $\delta_b=\delta$, and using $\delta_v^{(1)}(\boldsymbol{k})=\mu^2\delta_b^{(1)}(\boldsymbol{k})$,  we have \citep{Munoz2015}
\begin{equation}
B_{21}^{\rm G}=2(\alpha+\bar{T}_{21}\mu_1^{2})(
\alpha+\bar{T}_{21}\mu_2^{2})
 \left[ \gamma F(\boldsymbol{k_1},\boldsymbol{k_2})+\bar T_{21}\mu_3^2 G(\boldsymbol{k_1},\boldsymbol{k_2})\right] P_1P_2+\text{2 perm.},
\end{equation}
where $P _ { i } \equiv P _ { \delta } \left( \boldsymbol { k } _ { i } \right)$ is the power spectrum of the linear overdensity, and the ``2 perm.'' in the equation denotes permutations of indices (1,2) by (2,3) and (3,1). 
For a CDM-only universe, the mode coupling kernels $F(\boldsymbol{k_1},\boldsymbol{k_2})$ and $G(\boldsymbol{k_1},\boldsymbol{k_2})$ are \citep{Munoz2015}
\begin{eqnarray}
F \left( \boldsymbol{k_1}, \boldsymbol{k_2} \right) &=& \frac{5}{7}+  \frac{1}{2} \hat { k } _ { 1 } \cdot \hat { k } _ { 2 } \left( \frac { k _ { 1 } } { k _ { 2 } } + \frac { k _ { 2 } } { k _ { 1 } } \right) + \frac{2}{7} ( \hat { k } _ { 1 } \cdot \hat { k } _ { 2 } ) ^ { 2 }, \\
G \left( \boldsymbol{k_1}, \boldsymbol{k_2} \right) &=& \frac{3}{7} + \frac{1}{2} \ \hat { k } _ { 1 } \cdot \hat { k } _ { 2 } \left( \frac { k _ { 1 } } { k _ { 2 } } + \frac { k _ { 2 } } { k _ { 1 } } \right) + \frac{4}{7} ( \hat { k } _ { 1 } \cdot \hat { k } _ { 2 } ) ^ { 2 } .
\end{eqnarray}
where $\hat k_i=\boldsymbol k_i/|\boldsymbol k_i|$.

\emph{(iii)~Non-linear brightness-temperature dependency term $B_{21}^{\rm nl}$:}  The mapping from the density and velocity-gradient fields to the
21-cm brightness-temperature fluctuation is intrinsically nonlinear, which also contributes to the 21-cm bispectrum. This contribution can be written as \citep{Munoz2015}
\begin{equation}
B_{21}^{\rm nl}=(\alpha+\bar{T}_{21}\mu_1^{2})(\alpha+\bar{T}_{21}\mu_2^{2})\left[2\beta +\alpha(\mu_1^2 +\mu_2^2) +2\bar T_{21} \mu_1^2 \mu_2^2\right]P_1 P_2 +2\ \mathrm{perm.}.
\end{equation}

While the full information about three-point correlations is contained in the bispectrum $B_{21}(\mathbf{k}_1,\mathbf{k}_2,\mathbf{k}_3)$, its  amplitude is strongly modulated by the normalization of the power spectrum, the growth factor and (in our case) the mean 21-cm brightness temperature. This makes it less convenient for isolating the characteristic scale and shape dependence induced by primordial non-Gaussianity. Therefore, we use the reduced bispectrum, $Q_{21}$, which divides out the main dependence on the amplitude of power spectrum and emphasizes instead the configuration dependence of the three-point function  (e.g.  \cite{Scoccimarro2000,Bernardeau2001,Desjacques2010}).
For a given triangle configuration, $(\mathbf{k}_1,\mathbf{k}_2,\mathbf{k}_3)$ we define reduced bispectrum of 21-cm brightness-temperature fluctuation as 
\begin{equation}
Q_{21}(\boldsymbol{k}_1,\boldsymbol{k}_2,\boldsymbol{k}_3; z)
\equiv
\frac{B_{21}(\boldsymbol{k}_1,\boldsymbol{k}_2,\boldsymbol{k}_3; z)}
     {P_{21}(\boldsymbol{k_1}; z)\,P_{21}(\boldsymbol{k_2}; z) + P_{21}(\boldsymbol{k_2}; z)\,P_{21}(\boldsymbol{k_3}; z) + P_{21}(\boldsymbol{k_3}; z)\,P_{21}(\boldsymbol{k_1}; z)} ,
\label{eq:def_Q_general}
\end{equation}
where $B_{21}$ and $P_{21}$ are the bispectrum and power spectrum of the 21-cm brightness-temperature fluctuation field, respectively.
In this form, $Q_{21}$ is only weakly affected by a global rescaling of the field that multiplies both $B_{21}$ and $P_{21}$ by the same factor, but it remains highly sensitive to the shape and scale dependence generated by primordial non-Gaussianity and by non-linear gravitational evolution.

The spherically averaged 21-cm
reduced bispectrum is
\begin{equation}
    Q_{21}(k_1,k_2,k_3;z)=\frac{1}{4\pi}\int_0^{2\pi}\mathrm d \phi \int_{-1}^{1}\mathrm d \mu\   Q_{21}(k_1,k_2,k_3,\mu_1,\mu_2,\mu_3; z)
\end{equation}
Here, $(\mu,\phi)$ parameterize the direction of the line of sight relative to the $(k_1,k_2,k_3)$ triangle \citep{Yankelevich:2018uaz}. Therefore, the angular-average sums over all possible orientations of the triangles, 
and we can express $(\mu_1,\mu_2,\mu_3)$ as
\begin{equation}
    \mu_1=\mu,\quad  \mu_2=\mu_1\cos{\theta_{12}}+\sqrt{1-\mu_1^2} \sin{\theta_{12}}\sin{\phi},\quad \mu_3=-\frac{k_1}{k_3}\mu_1 -\frac{k_2}{k_3}\mu_2,
\end{equation}
where $\cos{\theta_{12}}=\hat{\boldsymbol{k}}_1\cdot\hat{\boldsymbol{k}}_2$ .

\subsection{The 21-cm signal in resonant and step models}
To quantify the impact of oscillatory primordial features on the 21-cm observables, we consider two representative templates: the resonant model which was first introduced in \cite{Chen2008} and later realized in axion-monodromy inflation \citep{McAllister2010,Flauger2010},
and the step model, in which a sharp feature in the inflation potential produces localized oscillatory features \citep{Adams2001,Chen2006,Chen2008}. In these two categories of models, the inflaton potential generates $k$-dependent oscillations in both the primordial power spectrum and the bispectrum.

\subsubsection{Resonant model}
The resonant model is predicted if there is small-scale structure in the inflationary potential \citep{Chen2008}.
One typical resonant model is realized in  
the axion-monodromy inflation, and   
the corresponding dimensionless primordial curvature power spectrum is \citep{Flauger2010,Xu2016}
\begin{equation}
\mathcal{P}^{\mathrm{res}}_{\Phi}(k) =
\mathcal{P}^{(0)}_{\Phi}(k)\,
\biggl[
    1 + \frac{8\,f_{\mathrm{res}}}{C_\omega^{2}}
    \cos\!\Bigl(C_\omega\ln\frac{k}{k_{\mathrm p}}\Bigr)
    \biggr],
    \label{PP_res}
\end{equation}
where $\mathcal{P}^{(0)}_{\Phi}(k)$ is the standard nearly scale-invariant dimensionless Bardeen-potential power spectrum, $f_{\rm res}$ is the amplitude of the resonant non-Gaussianity, and $C_{\omega} \equiv \omega /H_{\rm I}$ is the resonance ``frequency'' and $H_{\rm I}$ is the Hubble parameter during inflation. 
Models with higher resonance frequency $C_\omega$ is more suppressed by the factor proportional to
\(C_\omega^{-2}\).

The corresponding bispectrum is given by  \citep{Flauger2010,Xu2016}
\begin{equation}
\begin{aligned}
B^{\mathrm{res}}_{\Phi}
(\mathbf{k}_{1},\mathbf{k}_{2},\mathbf{k}_{3})
  &= \frac{80\pi^{4}}{3}\,
\frac{f_{\mathrm{res}}\,
\Delta_{\Phi}^{2}}
{k_{1}^{2}k_{2}^{2}k_{3}^{2}}
\Biggl[
\sin\!\Bigl(C_\omega\ln\frac{K}{k_{\mathrm p}}\Bigr)
+\frac{1}{C_\omega}   \cos\!\Bigl(C_\omega\ln\frac{K}{k_{\mathrm p}}\Bigr)
 \sum_{i\neq j}\frac{k_i}{k_j}
 +\mathcal{O}\!\bigl(C_\omega^{-2}\bigr)\Biggr]     ,
\end{aligned}
\label{Bphi_res}
\end{equation}
where $k_i=|\boldsymbol{k_i}|$, \(K\equiv k_{1}+k_{2}+k_{3}\) and
$\Delta_{\Phi}=\mathcal{P}_{\Phi}^{(0)}(k_{\rm p})$
is the amplitude of dimensionless Bardeen-potential power spectrum evaluated at $k_{\rm p}$.
The prefactor of \(C_\omega^{-1}\) implies that the cosine term dominates
for \(C_\omega\lesssim1\), and the sine term dominates the primordial  
bispectrum at higher resonance frequencies.

The PNG signal is strongest in the squeezed limit, i.e. $k_1\ll k_2 \simeq k_3$. 
In the squeezed limit, the scale ratio $r=k_1/k_2=k_1/k_3\ll1$, then 
the primordial bispectrum is simplified to 
\begin{equation}
\begin{aligned}
B^{\mathrm{res}}_{\Phi}
(\mathbf{k}_{1},\mathbf{k}_{2},\mathbf{k}_{3})
 &\approx 
\frac{80\pi^4f_\mathrm{res}\Delta_\Phi^2}{3} \cdot\frac{1}{r^2k_2^6}\cdot\left[\sin{\left(C_\omega\ln{\frac{2k_2}{k_p}}\right)}+\frac{2+2r}{rC_\omega}\cos{\left(C_\omega\ln{\frac{2k_2}{k_p}}\right)}\right]\\
&=F_\mathrm{const1}\cdot\frac{1}{r^2k_2^6}\cdot F_\omega,
\end{aligned}
\label{B1}
\end{equation}
where 
\begin{equation}
    F_\omega=
\sin{\left(C_\omega\ln{\frac{2k_2}{k_{\rm p}}}\right)}+\frac{2+2r}{rC_\omega}\cos{\left(C_\omega\ln{\frac{2k_2}{k_{\rm p}}}\right)},
\end{equation}
and
\begin{equation}
    F_{\rm const1}=\frac{80\pi^4f_\mathrm{res}\Delta_\Phi^2}{3}.
\end{equation}

Referring to Eq.~\eqref{B_I} and \eqref{Mkz}, the primordial contribution to the matter bispectrum in the squeezed limit is given by
\begin{equation}
\begin{aligned}
B_{\mathrm{m}}^{\rm I-res}
(\mathbf{k}_{1},\mathbf{k}_{2},\mathbf{k}_{3},z)
&= B^{\mathrm{res}}_{\Phi}(\mathbf{k}_{1},\mathbf{k}_{2},\mathbf{k}_{3})
M(k_1,z)M(k_2,z)M(k_3,z)\\
&=F_\mathrm{const1}\cdot F_\omega\cdot F_M^3 T(k_1)T^2(k_2)
\end{aligned}
\label{B2}
\end{equation}
where $F_M=2c^2D_+(z)/(3\Omega_m H_0^2)$. 
The scale ratio $r$ disappears from Eq.~\eqref{B2} except in $F_{\omega}$, because the factor $k_1^2k_2^2k_3^2$ included in $M(k_1,z)M(k_2,z)M(k_3,z)$ cancels out the same factor in the denominator in Eq.~(\ref{Bphi_res}), or equivalently the prefactor
$1/(r^2k_2^6)$ in Eq.\eqref{B1}. Under isotropic condition, the reduced matter bispectrum in the squeezed limit is then
\begin{equation}
\begin{aligned}
    Q_\mathrm{m}^\mathrm{I-res}(k_1,k_2,k_3,z)
    &=
    \frac{B_{\mathrm{m}}^{\rm I-res}(k_1,k_2,k_3,z)}{P_\mathrm{m}(k_1,z)P_\mathrm{m}(k_2,z)+2\mathrm{perm.}}\\
    &=\frac{B_{\mathrm{m}}^{\rm I-res}(k_1,k_2,k_3,z)}{2P_\mathrm{m}(k_1,z)P_\mathrm{m}(k_2,z)+P_\mathrm{m}^2(k_2,z)}
\end{aligned}
\label{Qres_squeezed_matter}
\end{equation}

Using the approximation that the scalar spectral index $n_s\approx 1$, we have
\begin{equation}
\begin{aligned}
    P_\mathrm{m}(k_i,z)P_\mathrm{m}(k_j,z)\approx A_mk_iT^2(k_i)\left(\frac{D_+(z)}{g(0)}\right)^2\times A_mk_jT^2(k_j)\left(\frac{D_+(z)}{g(0)}\right)^2
\end{aligned}
\label{B3}
\end{equation}
where $A_m=(8\pi^2/25)\cdot (\mathcal A_s /\Omega_m^2)\cdot \left[ c^4g^2(0)/(H_0^4 k_p^{n_s-1})\right]$ is the normalization amplitude of matter power spectrum, 
in which $g(0)$ denotes the present-day growth-suppression factor, so $D_+(z)/g(0)$ is the linear growth factor normalized to unity at $z=0$. We finally obtain
\begin{equation}
\begin{aligned}
Q_\mathrm{m}^\mathrm{I-res}(k_1,k_2,k_3,z)
    &\approx F_\mathrm{tot}\cdot\frac{T(k_1)}{k_2^2T^2(k_2) + 2k_1 k_2 T^2(k_1)}\cdot F_\omega,
\end{aligned}
\label{Qres_squeezed_matter2}
\end{equation}
with 
\begin{equation}
    F_\mathrm{tot}=\frac{10f_\mathrm{res}\Omega_mH _0^2}{c^2 D_+(z)}
\end{equation}
Here the explicit prefactor $1/r^2$ in the primordial bispectrum is
canceled by the $k^2$ factors in $M(k,z)$, while the squeezed-limit
dependence on $r$ remains through the oscillatory factor $F_\omega$ and
through the relation $k_1=rk_2$ in the denominator. Moreover, the
dominant term in the denominator depends on scale. For $k\ll k_{\rm eq}$,
$T(k)\simeq 1$ and the term proportional to $k_2^2T^2(k_2)$ dominates,
whereas on smaller scales the transfer-function suppression of $T(k_2)$
can make the cross term $2k_1k_2T^2(k_1)$ non-negligible.

For the resonant model, we can now compute the three-dimensional power spectrum and reduced bispectrum in redshift space with the redshift space distortion (RSD) effect included. To illustrate the relative contribution of different components and the sensitivities to various parameters, we first present the observables in their spherical-averaged form.

\begin{figure}[htbp]
  \centering
  \includegraphics[width=0.7\textwidth]{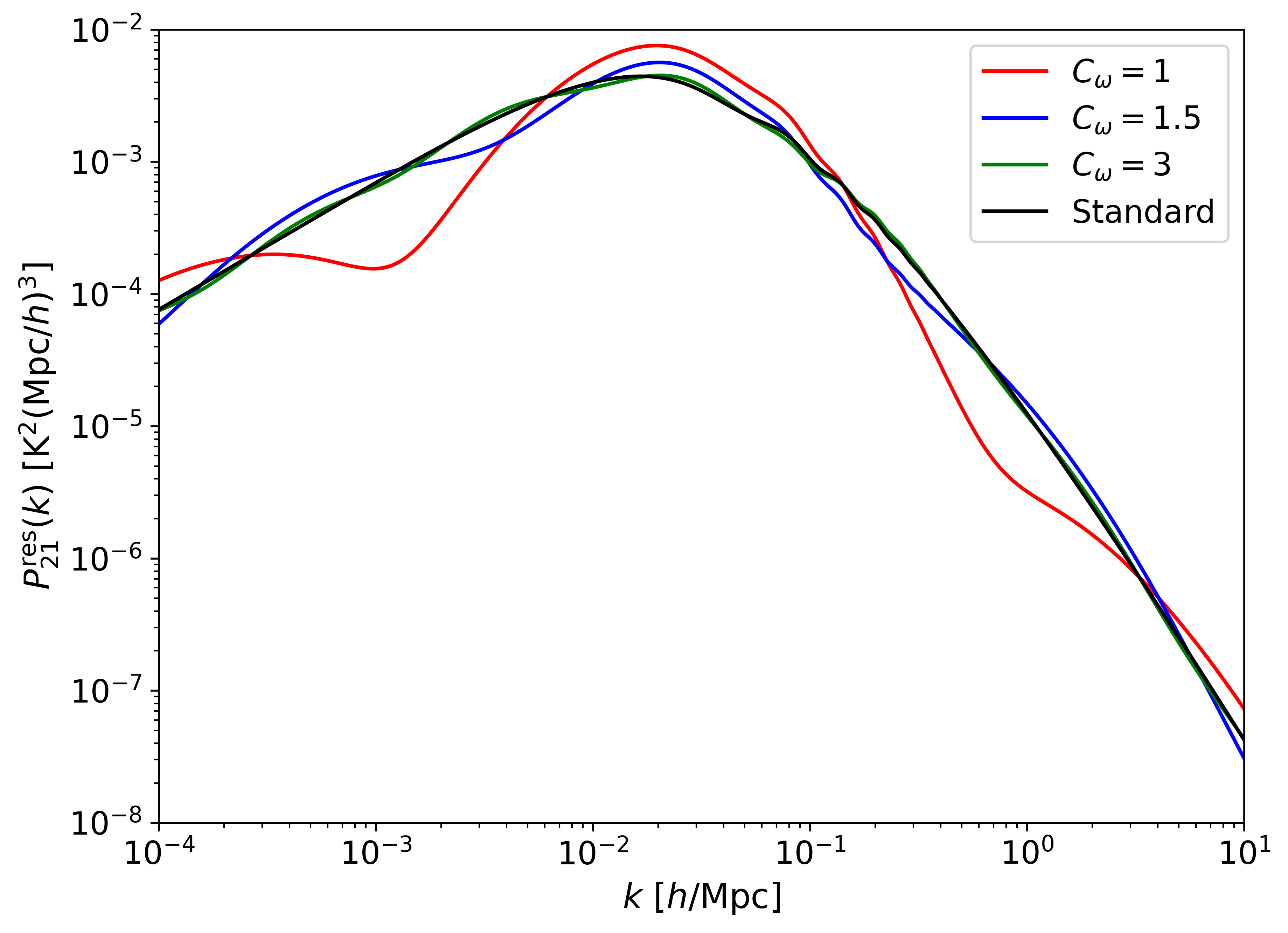}
  \caption{The
  21-cm brightness-temperature power spectrum $P_{21}^{\rm res}(k)$ for the resonant  
  model at $z=30$, 
  for several values of the oscillation frequency $C_\omega$. Here $f_{\rm res}=0.1$ is adopted. The red, blue, and green curves correspond to $C_\omega=1$, $1.5$, and $3$, respectively, and the black curve denotes the standard case.
  }
  \label{fig:resonant_P}
\end{figure}

Fig.~\ref{fig:resonant_P} illustrates the 21-cm brightness-temperature power spectrum $P_{21}(k)$ in the resonant 
model at $z=30$ assuming $f_{\rm res} = 0.1$, for the oscillation frequencies $C_\omega=1, 1.5$, and 3, respectively. Compared with the standard case with no feature, the resonant model induces oscillations  in the power spectrum. 
While a larger $C_\omega$ implies a higher oscillation frequency, it also suppresses the resonance amplitude by the factor ${C_\omega^{-2}}$ (as in Eq.\eqref{PP_res}), so the overall deviation of $P_{21}(k)$ from the standard model becomes weaker.

\begin{figure}[htbp]
  \centering
  \includegraphics[width=\textwidth]{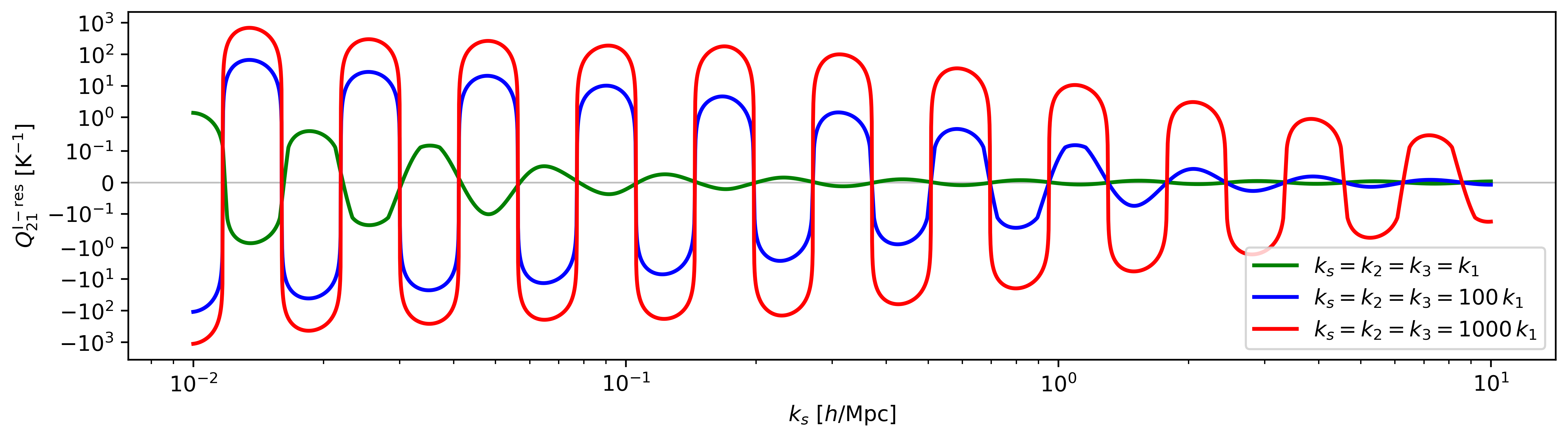}
\caption{The primordial contribution to the
reduced 21-cm brightness-temperature bispectrum  $Q_{21}^{\rm I-res}(k_1,k_2,k_3)$ for the resonant  
model at $z=30$.
We fix $f_{\rm res}=0.01$ and $C_\omega=10$, and plot $Q_{21}^{\rm I-res}$ as a function of the characteristic scale $k_s \equiv k_2=k_3$ for triangle families with $k_1/k_s=1,10^{-2}$, and $10^{-3}$ (from equilateral to squeezed), respectively, using colored lines as indicated in the legend. 
    }
\label{fig:resonant_Q_1}
\end{figure}

\begin{figure}[htbp]
  \centering
  \includegraphics[width=\textwidth]{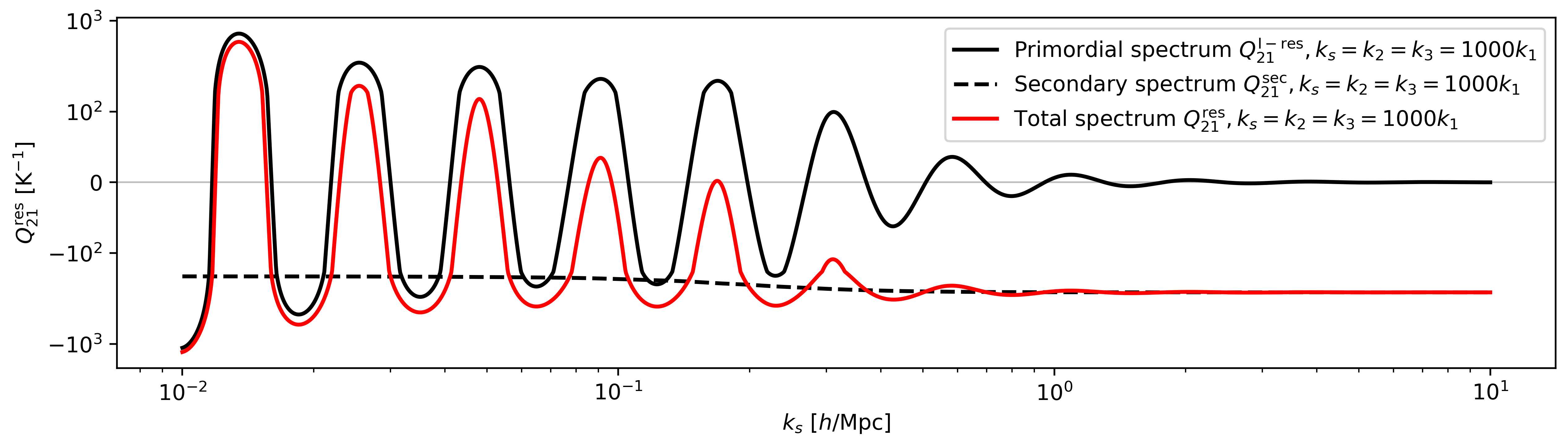}
  \caption{
Comparison of the primordial and secondary contributions to the
reduced 21-cm brightness-temperature bispectrum in the resonant feature model, for a strongly squeezed configuration with  $k_s\equiv k_2=k_3=1000\,k_1$ at $z=30$, assuming $C_\omega=10,\  f_\mathrm{res}=0.01$.
 The solid black curve shows the primordial contribution $Q_{21}^{\rm I-res}$, the black dashed line shows the secondary contribution $Q_{21}^{\rm sec}=Q_{21}^{\rm G}+Q_{21}^{\rm nl}$, and the red curve shows the total reduced bispectrum $Q_{21}^{\rm res}=Q_{21}^{\rm I-res}+Q_{21}^{\rm sec}$.
 The secondary term is nearly scale-independent and negative, and dominates the total signal at large $k_s$. 
    }
    \label{fig:resonant_sec}
\end{figure}

\begin{figure}[htbp]
  \centering
  \includegraphics[width=\textwidth]{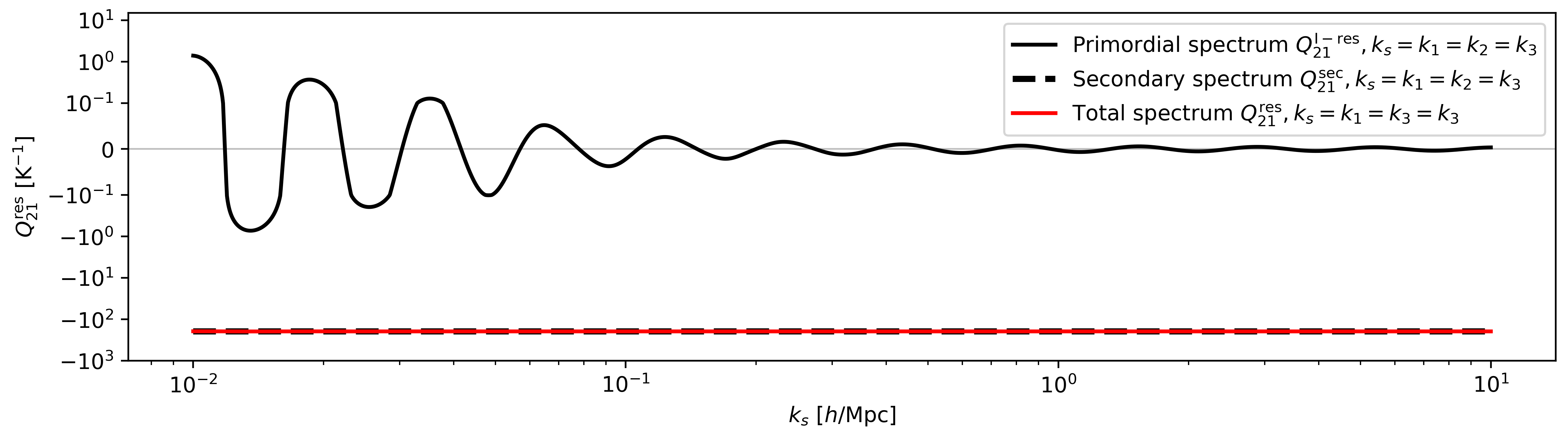}
  \caption{
Comparison of the primordial and secondary contributions to the
reduced 21-cm brightness-temperature bispectrum in the resonant feature model for an equilateral configuration with  $k_1=k_2=k_3\equiv k_s$ at $z=30$, assuming $C_\omega=10,\ f_\mathrm{res}=0.01$.
  The solid black curve shows the primordial contribution $Q_{21}^{\rm I-res}$, the black dashed line shows the secondary contribution $Q_{21}^{\rm sec}$, and the red curve shows the total reduced bispectrum $Q_{21}^{\rm res}=Q_{21}^{\rm I-res}+Q_{21}^{\rm sec}$.
Unlike the strongly squeezed case, the secondary contribution dominates the total signal throughout the scales for the equilateral configuration.
  }
    \label{fig:resonant_sec_eq}
\end{figure}

Assuming $f_{\rm res} = 0.01$,
Fig.~\ref{fig:resonant_Q_1} shows the primordial contribution to the reduced 21-cm brightness-temperature bispectrum $Q_{21}(k_1,k_2,k_3)$ in the resonant model, for the equilateral configuration ($k_1=k_2=k_3$) and two cases of squeezed configurations, i.e. $k_1\ll k_2 \simeq k_3$. In the resonant model, the 21-cm bispectrum shows significant log-periodic modulations in the characteristic scale $k_s\equiv k_2 = k_3$. Note that we have adopted logarithmic vertical axis to show the huge difference in the modulation amplitude. 
The more squeezed configuration has larger amplitude of modulation, and therefore, we expect higher constraining power from squeezed configurations.

In Fig.~\ref{fig:resonant_sec}, we compare the primordial contribution (black solid line), secondary contribution (black dashed line), and the total reduced bispectrum (red curve) for the strongly squeezed configuration with $k_s\equiv k_2=k_3=1000\,k_1$. 
The secondary term, arising from nonlinear gravitational evolution and the nonlinear relation between 21-cm brightness and matter field, is nearly independent of the primordial model 
parameters and remains roughly a constant, which is negative over the range of $k_s$ considered. On the other hand, the primordial contribution is strongly scale-dependent and oscillatory. On large scales (small $k_s$) the primordial 
term dominates and $Q_{21}^{\rm res}$ closely follows $Q_{21}^{\rm I-res}$, and at larger $k_s$  
the total signal becomes increasingly dominated by the secondary term, so that $Q_{21}^{\rm res}$ approaches the nearly flat negative value of the secondary contribution.
For comparison, we also show the case for equilateral configuration in Fig.~\ref{fig:resonant_sec_eq}. Unlike the strongly squeezed case, the secondary contribution always dominants the total signal over the plotted scale range in the equilateral configuration, so the total reduced bispectrum hardly shows any oscillatory features and is largely determined by the secondary bispectrum $Q_{21}^{\rm sec}$.

\begin{figure}[htbp]
  \centering
  \includegraphics[width=\textwidth]{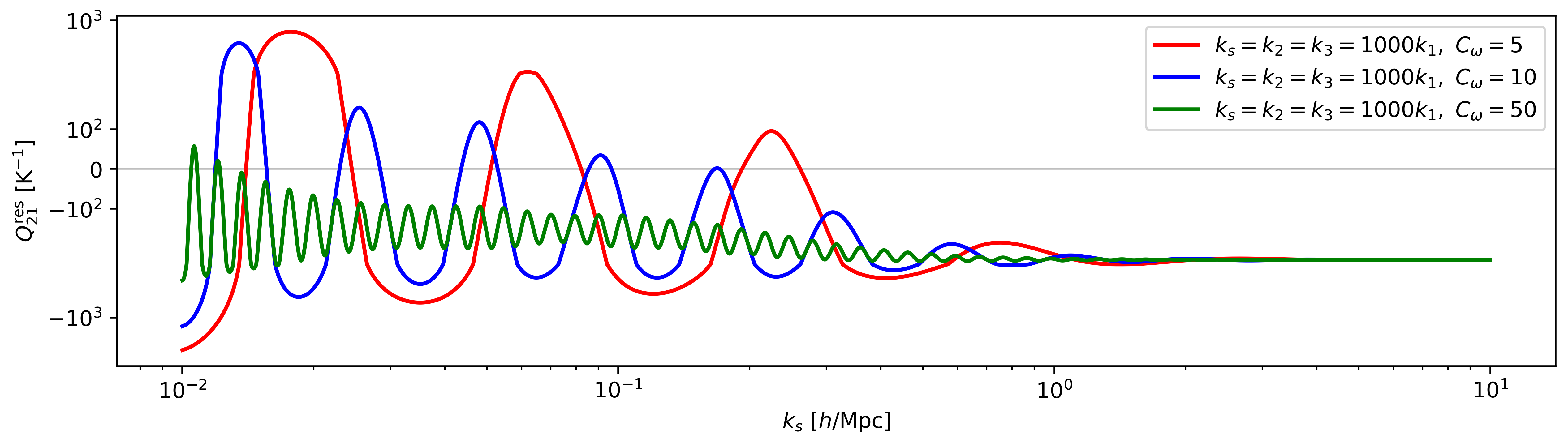}
  \caption{
  Total reduced 21-cm brightness-temperature bispectrum $Q_{21}^{\rm res}(k_1,k_2,k_3)$ (primordial + secondary) for the resonant PNG model at $z=30$. We fix $f_{\rm res}=0.01$ and plot $Q_{21}^{\rm res}$ as a function of $k_s \equiv k_2=k_3$ for triangle families with fixed $k_1/k_s=10^{-3}$, for different resonant frequencies of $C_\omega=5$, 10,  and 50 (with colors indicated in the legend), respectively. 
    }
    \label{fig:resonant_Q_2}
\end{figure}

As the primordial term 
contributes most in the squeezed limit, we plot the total reduced 21-cm bispectrum for a squeezed configuration in
Fig.~\ref{fig:resonant_Q_2} for various 
resonant frequency
parameters $C_\omega$ at a fixed $f_{\rm res} = 0.01$.  Changing $C_\omega$ primarily alters the oscillatory frequency in $\ln k_s$, while a higher frequency parameter also suppresses the oscillatory amplitude.
Combining Fig.~\ref{fig:resonant_sec} and Fig.~\ref{fig:resonant_Q_2}, it is found that for small $k_s$ or large scales, the oscillatory feature 
contribution can dominate $Q_{21}^{\rm res}$ and will be visible in the total signal, whereas
for sufficiently large $C_\omega$ and large $k_s$ the total reduced bispectrum
is dominated by the secondary non-Gaussianities. 
Note that the secondary contributions to the bispectrum also depends on the power spectrum $P_{\delta}(k)$, and therefore, they inherit some sensitivity to the feature parameters through the oscillatory features in $P_{\delta}(k)$.
Thus, even when the direct primordial 
contribution to $Q_{21}$ is weak and the signal-to-noise ratio is low, the secondary and total reduced
bispectra can still carry some useful information about the underlying oscillatory feature.

\subsubsection{Step Model}

Step model of inflation has a sharp, localized feature in the inflaton potential at the conformal time $\tau_f$, which produces oscillatory features in the primordial power spectrum and bispectrum \citep{Adams2001,Adshead2012}. In leading order, the  dimensionless primordial power spectrum is modified as \citep{Adshead2012}
\begin{equation}
\mathcal{P}^{\mathrm{step}}_{\Phi}(k)
=\mathcal{P}^{(0)}_{\Phi}(k)\,
\exp\!\left[
-\frac{2}{3}\,\epsilon_{\mathrm{step}}\,
W^{'}(k\tau_f)\,
\mathcal D\!\left(\frac{k\tau_f}{\beta_{\rm step}}\right)
\right],
\end{equation}
with
\begin{equation}
W^{'}(x)\equiv\left(-3+\frac{9}{x^{2}}\right)\!\cos(2x)
+\left(15-\frac{9}{x^{2}}\right)\!\frac{\sin(2x)}{2x},
\qquad
\mathcal D(y)\equiv\frac{\pi y}{\sinh(\pi y)}.
\label{Dy}
\end{equation}
Here,  $\epsilon_{\mathrm{step}}\ll 1$ is the height of step in the potential, $\beta_{\rm step}\gg 1$ controls the sharpness of  the step (a larger $\beta_{\rm step}$ gives a broader oscillatory feature and more slowly-damping envelope), and $\tau_f$ sets the scale and frequency of the oscillations. 

In the step model, the primordial  bispectrum can be written as \citep{Chen2006}.
\begin{equation}
\begin{aligned}
B^{\mathrm{step}}_{\Phi}(\mathbf{k}_{1},\mathbf{k}_{2},\mathbf{k}_{3})
&= \frac{5}{12}\,\epsilon_{\mathrm{step}}\,
\mathcal D\!\left(\frac{K\tau_f}{2\beta_{\rm step}}\right)\,
\frac{(2\pi)^{4}\,\Delta_{\Phi}^{2}}{k_{1}^{2}k_{2}^{2}k_{3}^{2}}
\\
&\quad\times
\bigg\{
\left[
\frac{k_{1}^{2}+k_{2}^{2}+k_{3}^{2}}{k_{1}k_{2}k_{3}\,\tau_f}
- K\tau_f
\right] K\tau_f \cos(K\tau_f)
\\
&\qquad\qquad
-\left[
\frac{k_{1}^{2}+k_{2}^{2}+k_{3}^{2}}{k_{1}k_{2}k_{3}\,\tau_f}
- \frac{\sum_{i\neq j}k_{i}^{2}k_{j}}{k_{1}k_{2}k_{3}}\,
K\tau_f
\right]\sin(K\tau_f)
\bigg\},
\end{aligned}
\label{Bphi_step}
\end{equation}
where $k_{i}\equiv|\mathbf{k}_{i}|$, $K\equiv k_{1}+k_{2}+k_{3}$, and $\Delta_{\Phi}\equiv\mathcal{P}^{(0)}_{\Phi}(k_{\mathrm p})$.

We can also derive an analytical expression of the primordial contribution to the reduced matter  bispectrum, and also the reduced 21-cm bispectrum using Eq.~(\ref{B_I}) and (\ref{eq:def_Q_general}), in the squeezed limit.
From Eq.~\eqref{Bphi_step}, defining $x\equiv 2k_2\tau_f$, we have
\begin{equation}
\begin{aligned}
    B^{\mathrm{step}}_{\Phi}(\mathbf{k}_{1},\mathbf{k}_{2},\mathbf{k}_{3})
&\approx 
\frac{5}{12}\,\epsilon_{\mathrm{step}}\,
\mathcal D\!\left(\frac{x}{2\beta_{\rm step}}\right)\,
\frac{(2\pi)^{4}\,\Delta_{\Phi}^{2}}{k_{1}^{2}k_{2}^4}\,\frac{1}{r}\, \mathcal J(x)
\end{aligned}
\end{equation}
where the oscillation term $\mathcal J(x)$ is
\begin{equation}
\mathcal J(x)=4\cos{x}+\left(2x-\frac{4}{x}\right)\sin{x}.
\end{equation}
Similar to the derivation in the case of resonance model, the primordial term of the matter bispectrum in the squeezed limit is 
\begin{equation}
\begin{aligned}
Q_\mathrm{m}^\mathrm{I-step}(k_1,k_2,k_3,z)
&=F_Q\cdot\frac{T(k_1)}{k_2^2T^2(k_2) + 2k_1 k_2 T^2(k_1)} \, \mathcal D\left(\frac{x}{2\beta_{\rm step}}\right)\, \frac{\mathcal J(x)}{r}
\end{aligned}
\end{equation}
where $r=k_1/k_2=k_1/k_3\ll1$, $F_Q$ is a redshift-dependent constant, i.e. 
\begin{equation}
F_Q=\frac{5\,\epsilon_\mathrm{step}\Omega_mH_0^2}{2\,c^2D_+(z)},
\end{equation}
and $\mathcal D\left(\frac{x}{2\beta}\right)$ is the damping function given in Eq.~\eqref{Dy}.

\begin{figure}[htbp]
  \centering
  \includegraphics[width=0.7\textwidth]{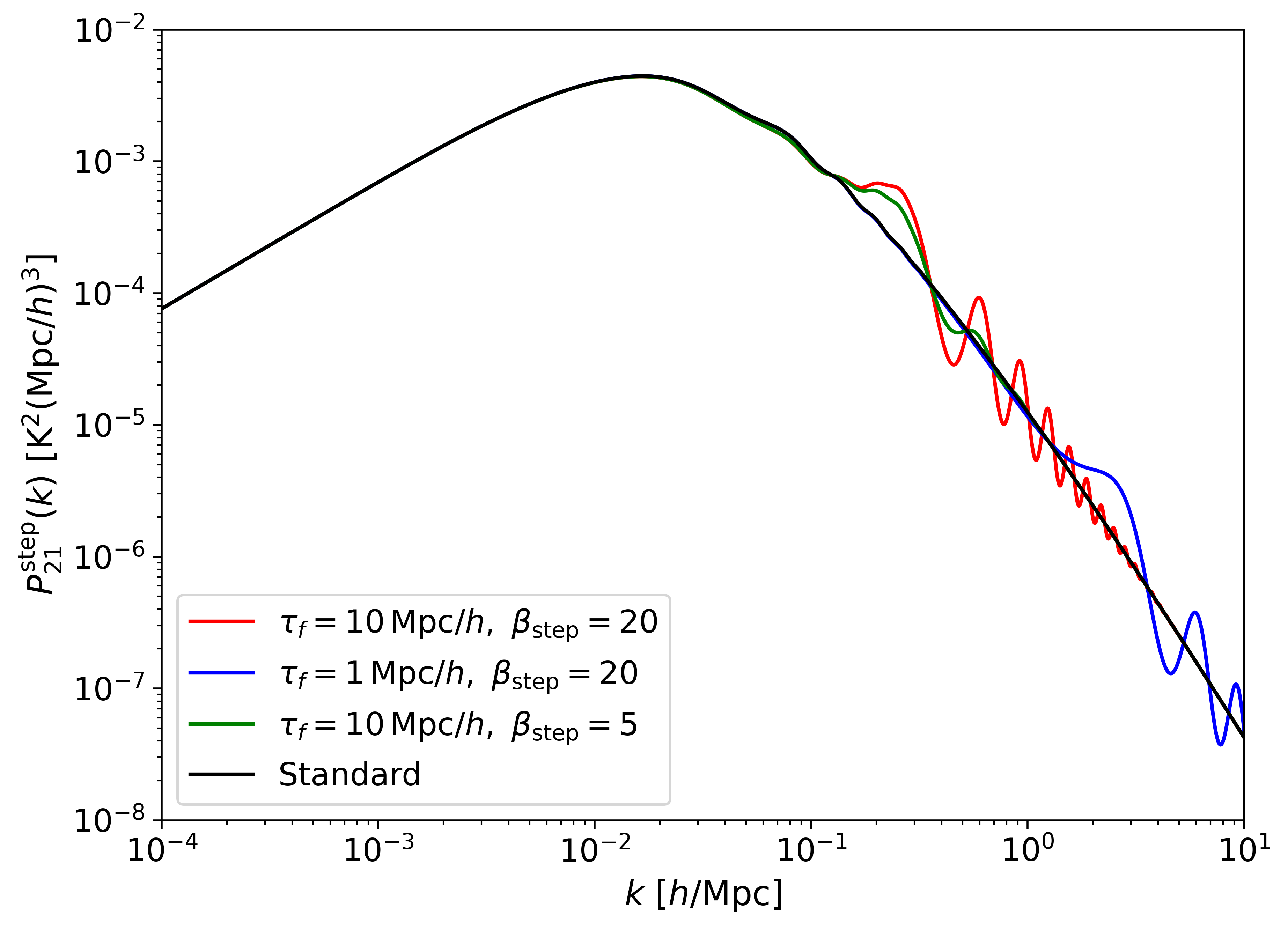}
  \caption{
 The 21-cm brightness-temperature power spectrum $P_{21}^{\rm step}(k)$ for the step 
 model at $z=30$, plotted as a function of $k$, for several combinations of parameters $(\tau_f,\beta_\mathrm{step})$ as indicated in the legend. Here $\epsilon_{\rm step}=0.5$ is adopted. 
 The black curve shows the standard case.
  }
    \label{fig:step_P}
\end{figure}

For the step model, 
we also present the 21-cm brightness-temperature power spectrum and reduced bispectrum in the spherical-averaged form, to illustrate the different contributions and effects of various parameters.
Fig.~\ref{fig:step_P} shows the 21-cm  power spectrum $P_{21}(k)$ in the step 
model at $z=30$ for several choices of $\tau_f$ and $\beta_{\rm step}$, together with the standard case. Different from the resonant model, the step model generates a localized oscillatory feature in the power spectrum. The parameter $\tau_f$ mainly determines the location of this feature in $k$-space, while $\beta_{\rm step}$ controls how broadly the oscillation extends over the $k$ range.

\begin{figure}[htbp]
  \centering
  \includegraphics[width=\textwidth]{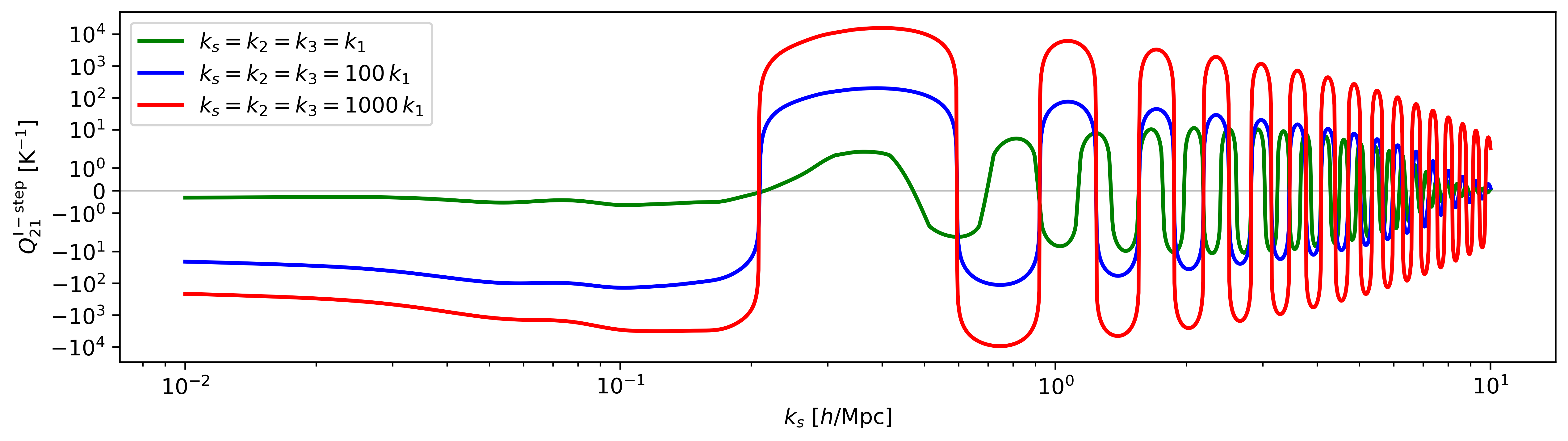}
  \caption{
The primordial contribution to the
reduced 21-cm brightness-temperature bispectrum $Q_{21}^{\rm I-step}(k_1,k_2,k_3)$ for the step  
model at $z=30$. 
We fix $\epsilon_{\rm step}=0.01$, $\tau_f=5\,{\rm Mpc}/h$, and $\beta_{\rm step}=20$, and plot $Q_{21}^{\rm I-step}$ as a function of the characteristic scale $k_s \equiv k_2=k_3$ for triangle families with $k_1/k_s=1,10^{-2}$, and $10^{-3}$ (from equilateral to squeezed), respectively, using colors as indicated in the legend. 
    }
    \label{fig:step_Q_1}
\end{figure}

In Fig.~\ref{fig:step_Q_1}, we first show the PNG term of the bispectrum for several triangle configurations in the step model, from the equilateral case, to the squeezed cases at fixed parameters $(\epsilon_{\rm step},\tau_f,\beta_{\rm step})=(0.01,5\,{\rm Mpc}/h,20)$. 
Similar to the case of resonance model, here we also see that the more squeezed cases have larger amplitudes.

\begin{figure}[htbp]
  \centering
  \includegraphics[width=\textwidth]{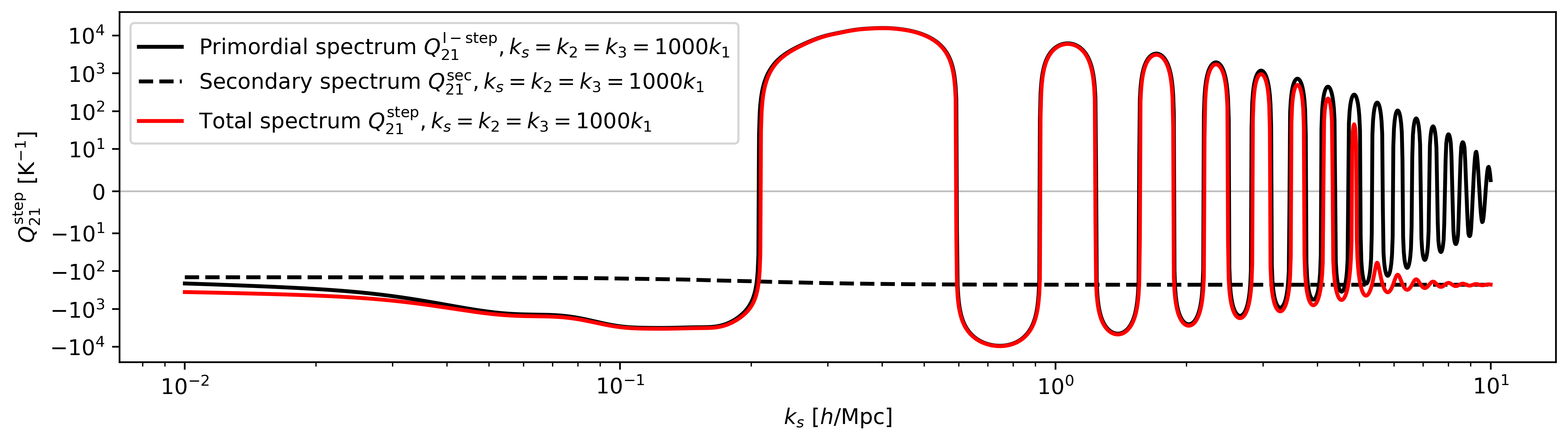}
  \caption{
Comparison of the primordial and secondary contributions to the
reduced 21-cm brightness-temperature bispectrum in the step feature model for a strongly squeezed configuration with $k_s\equiv k_2=k_3=1000\,k_1$ at $z=30$, assuming $\tau_f=5\, \mathrm{Mpc}/h,\ \beta=20$, and $\epsilon_\mathrm{step}=0.01$.
The solid black curve shows the primordial contribution $Q_{21}^{\rm I-step}$, the black dashed line shows the secondary contribution $Q_{21}^{\rm sec}$, and the red curve shows the total reduced bispectrum $Q_{21}^{\rm step}=Q_{21}^{\rm I-step}+Q_{21}^{\rm sec}$.
    }
    \label{fig:step_sec}
\end{figure}

\begin{figure}[htbp]
  \centering
  \includegraphics[width=\textwidth]{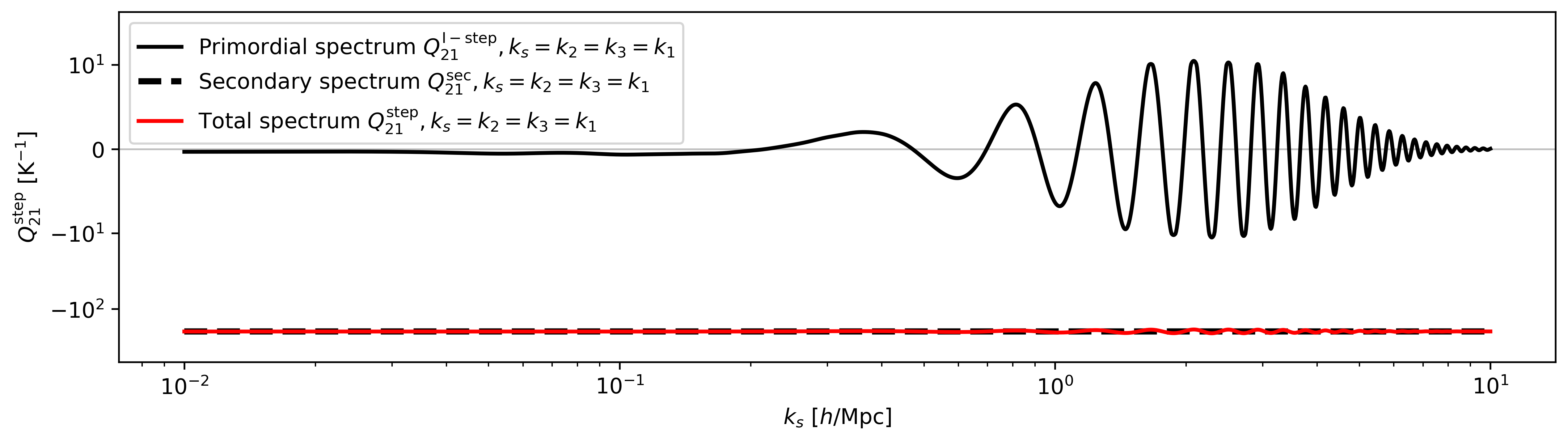}
  \caption{
Comparison of the primordial and secondary contributions to the
reduced 21-cm brightness-temperature bispectrum in the step feature model for an equilateral configuration with $k_s \equiv k_1=k_2=k_3$ at $z=30$, assuming $\tau_f=5\, \mathrm{Mpc}/h$, $\beta=20$, and $\epsilon_\mathrm{step}=0.01$.
The solid black curve shows the primordial contribution $Q_{21}^{\rm I-step}$, the black dashed line shows the secondary contribution $Q_{21}^{\rm sec}$, and the red curve shows the total reduced bispectrum $Q_{21}^{\rm step}=Q_{21}^{\rm I-step}+Q_{21}^{\rm sec}$.
Unlike the strongly squeezed case, the secondary term remains the dominant contribution to the total signal throughout the scales for the equilateral configuration. 
    }
    \label{fig:step_sec_equilateral}
\end{figure}

\begin{figure}[htbp]
  \centering
  \includegraphics[width=\textwidth]{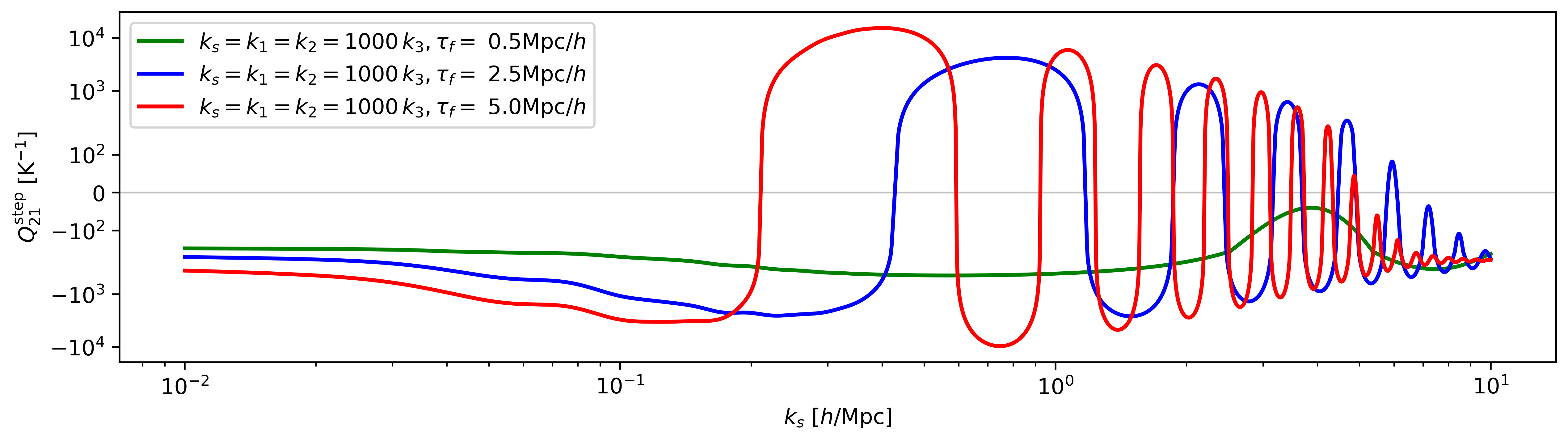}
  \caption{
 The  total reduced 21-cm brightness-temperature bispectrum $Q_{21}^{\rm step}(k_1,k_2,k_3)$ (primordial + secondary) for the step feature 
 model at $z=30$, with fixed $\epsilon_{\rm step}=0.01$ and $\beta_{\rm step}=20$ and varying transition scales of $\tau_f=0.5$, 2.5, and 5.0 ${\rm Mpc}/h$, respectively, showing with colors as indicated in the legend. We plot $Q_{21}^{\rm step}$ as a function of $k_s \equiv k_2=k_3$ for triangle families in the squeezed limit with fixed $k_1/k_s=10^{-3}$.
    }
    \label{fig:step_Q_2}
\end{figure}

\begin{figure}[htbp]
  \centering
  \includegraphics[width=\textwidth]{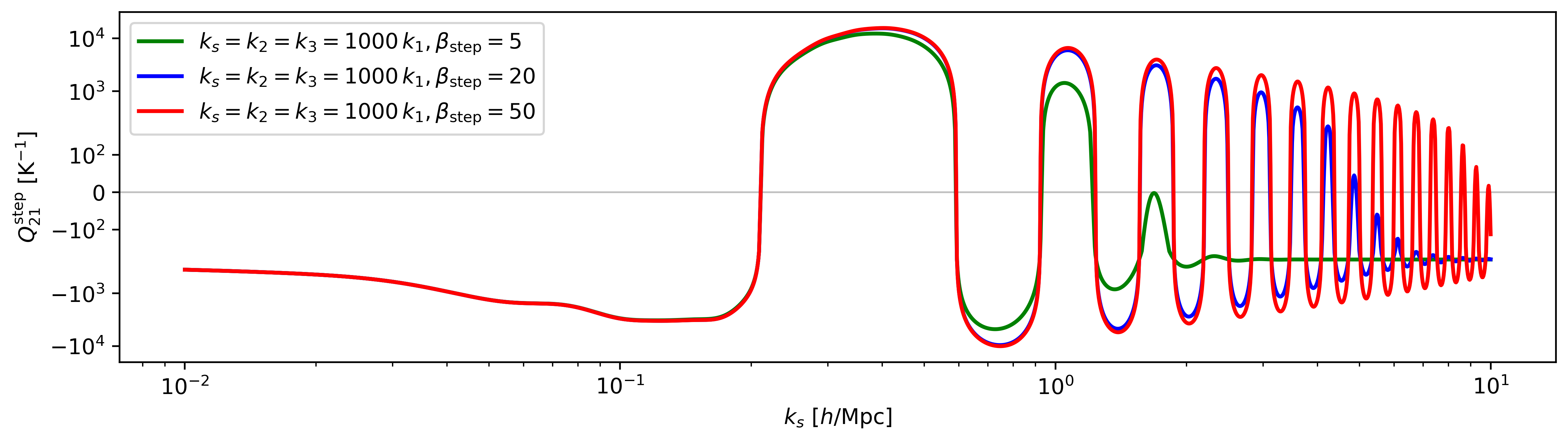}
  \caption{
The total reduced 21-cm brightness-temperature bispectrum $Q_{21}^{\rm step}(k_1,k_2,k_3)$ (primordial + secondary) for the step feature 
model at $z=30$, with fixed $\epsilon_{\rm step}=0.01$ and $\tau_f=5\,{\rm Mpc}/h$ and varying steepness parameters of $\beta_{\rm step}=5$, 20, and 50, respectively, showing with colors as indicated in the legend. We plot $Q_{21}^{\rm step}$ as a function of $k_s \equiv k_2=k_3$ for triangle families in the squeezed limit with fixed $k_1/k_s=10^{-3}$. 
    }
    \label{fig:step_Q_3}
\end{figure}

In Fig.~\ref{fig:step_sec}, we show the primordial contribution (black solid line) and the secondary contribution (black dashed line) to the total reduced bispectrum (red curve) for the strongly squeezed configuration with $k_s=k_1=k_2=1000\,k_3$. The secondary term is nearly the same as in the resonant feature model, arising from nonlinear gravitational evolution and the nonlinear relation between 21-cm brightness temperature and baryon density, and is nearly scale independent and negative over the range of $k_s$ considered. By contrast, the primordial contribution in the step feature model is highly scale dependent. At small $k_s$ the total signal closely follows $Q_{21}^{\rm I-step}$, while at larger $k_s$ the secondary and primordial contributions become comparable, and the total signal is eventually dominated by the secondary term, so that $Q_{21}^{\rm step}$ approaches the nearly flat, negative secondary contribution.

For comparison, we also show the equilateral configuration in Fig.~\ref{fig:step_sec_equilateral}. Unlike the strongly squeezed case, the secondary term remains dominant over the plotted scale range in the equilateral configuration, while the primordial contribution only appears as a relatively small oscillatory correction. As a result, the total reduced bispectrum is largely determined by $Q_{21}^{\rm sec}$ rather than the primordial contribution.

In Figs.~\ref{fig:step_Q_2} and \ref{fig:step_Q_3}, we vary the parameters $\tau_f$ and $\beta_{\rm step}$ in the step feature model, respectively. At fixed $\epsilon_{\rm step}$ and $\beta_{\rm step}$, varying the parameter $\tau_f$ 
shifts the location and phase of the feature approximately according to a $\tau_f^{-1}$ scaling. 
At fixed $\epsilon_{\rm step}$ and $\tau_f=5\,{\rm Mpc}/h$, increasing the parameter $\beta_{\rm step}$
makes the transition sharper and the high-$k_s$ oscillations less damped, producing more  oscillatory signatures potentially visible if structures on relevant scales can be measured.
Together, these figures show that $\tau_f$ sets \emph{where} the feature appears in, while $\beta_{\rm step}$ sets how \emph{sharp} and \emph{persistent} it is.

\section{The Forecast} 

We quantify the measurement 
uncertainties on the amplitude parameters $f_{\mathrm{res}}$ in the resonant model and  
$\epsilon_{\mathrm{step}}$ in the step model through a Fisher‐matrix analysis constructed for both the 21-cm power spectrum and the reduced 21-cm bispectrum from the Dark Ages.

\subsection{The Fisher formalism}
\label{sec:fisher_formalism}

In 21-cm intensity mapping experiments, the foreground contaminates a wedge region in the $( k_{\perp},k_{\parallel})$ space due to intrinsic chromatic response of an interferometer, where $k_{\perp}$ and $k_{\parallel}$ are the Fourier modes  perpendicular and parallel with the line of sight respectively 
\citep{Morales:2012kf,Parsons:2012qh,Liu:2014yxa}. 
The wedge edge is set by $k_{\parallel}\leq m(z)\, k_{\perp}$, in which $m(z)$ only depends on redshift.
Previous forecasts usually take an approximated approach by excluding a fraction of modes potentially contaminated by foregrounds (e.g. \cite{Xu2016}), or model the residual foregrounds as a contribution to the power spectrum (e.g. \cite{Bull:2014rha}). Then one can use the standard Fisher formalism.

The standard Fisher matrix for the power spectrum takes the form of 
\begin{equation}    F_{\alpha\beta}^P\equiv\sum_{k_{min}}^{k_{\max }}\frac{\partial P_{21}}{\partial \alpha }\frac{\partial P_{21}}{\partial\beta}\frac{1}{\Delta P^{2}},
\end{equation}
where $P_{21}$ means the signal power spectrum excluding noise, and the variance of the isotropic power spectrum is
\begin{equation}
    \Delta P ^2=\frac{P_{\rm tot}^2(k)}{N_P},
\end{equation}
where 
\begin{equation}
N_P=(4\pi k^2 \Delta k /V_f/2)\times(V_{\rm survey}/V_{\rm one}),
\end{equation}
and $P_{\rm tot}(k)=P_{21}(k) + P_N(k)$ in which $P_N$ is the noise power spectrum. Here $V_{\rm survey}$ is the total survey volume of each redshift bin, 
which can be split into multiple pointings with individual cubic observing volumes of $V_{\rm one}$ that are determined by the field of view (FOV) of the telescope. For each such volume, the fundamental wavenumber is $k_f = (2\pi) / \sqrt[3]{V_{\rm one}}$, so that the volume of a fundamental Fourier cell is $V_f = k_f^3 = (2\pi)^3 / V_{\rm one}$. 
Here we take the minimum accessible wavenumber to be $k_{\min} = k_f$ \citep{Chiang:2017vsq,Chen:2024exy}.
The Fisher matrix from the reduced 21-cm bispectrum is 
\begin{equation}
F_{\alpha\beta}^Q\equiv\sum_{k_{1}=k_{\min}}^{k_{\max }}\sum_{k_{2}=k_{\min}}^{k_{1}}\sum_{k_{3}=k_{\min}^{\star}}^{k_2}\frac{\partial Q_{21}}{\partial \alpha }\frac{\partial Q_{21}}{\partial\beta}\frac{1}{\Delta Q_{21}^{2}},
\end{equation}
where the three sums are over all combinations of $\vec{k_1}$, $\vec{k_2}$, and $\vec{k_3}$ that form triangles, with $k_{\rm min}^{*}=\max{(k_{\rm min},|k_1 - k_2|)}$. The variance of the isotropic reduced bispectrum is \citep{Xu2016}
\begin{equation}
    \Delta Q _ { 21 } ^ { 2 } \left( k _ { 1 } , k _ { 2 } , k _ { 3 } \right) \simeq \frac { \Delta B _ { 21 } ^ { 2 } \left( k _ { 1 } , k _ { 2 } , k _ { 3 } \right) } { \left[ P _ { 21 } \left( k _ { 1 } \right) P _ { 21 } \left( k _ { 2 } \right) + ( 2\, perm. ) \right]^2 }.
\end{equation}
and the variance of isotropic bispectrum is given by \citep{Sefusatti2006},
\begin{equation}
    \Delta B _ { 21 } ^ { 2 } \left( k _ { 1 } , k _ { 2 } , k _ { 3 } \right) \simeq \frac{ s_{ 1 2 3 } \left[(2\pi)^3/V _ { \mathrm { f }}\right]    P _ { \mathrm { t o t } } \left( k _ { 1 } \right) P _ { \mathrm { t o t } } \left( k _ { 2 } \right)P _ { \mathrm { t o t } } \left( k _ { 3 } \right)}{N_B},
\end{equation}
where  $N_B=(V_B/k_f^6)\,(V_{\rm survey}/V_{\rm one})$, $V_B=8\pi^2k_1k_2k_3(\Delta k)^3$ and $s_{123}=6,2,1$ for equilateral, isosceles, and general triangles, respectively. 
The integral form at each redshift bin is then \citep{Bull2024} 
\begin{equation}
\begin{aligned}
F_{\alpha \beta}^{Q} = \frac{V_{\rm survey}}{4\pi^{2}} \frac{1}{6} 
\int_{-1}^{1} d\mu_{1} \int_0^{2\pi} d\phi \int_{k_{\min}}^{k_{\max}} & dk_{1} dk_{2} dk_{3}  \left\{  \frac{(k_{1} k_{2} k_{3})\left[ P _ { 21 } \left( k _ { 1 } \right) P _ { 21 } \left( k _ { 2 } \right) + ( 2 \mathrm{pe rm.} ) \right]^2}{(2\pi)^3}\right.
    \\
& \left.\quad \times \frac{\partial_{\alpha}Q_{21}(z_{i})\,\partial_{\beta}Q_{21}(z_{i})}
    {P_{\rm tot}(k_{1},\mu_{1},z_{i})\,P_{\rm tot}(k_{2},\mu_{2},z_{i})\,P_{\rm tot}(k_{3},\mu_{3},z_{i})}\right\},
\end{aligned}
\end{equation}
where $\mu_i=k_\parallel/|\boldsymbol{k}|$, and $z_i$ is the central redshift of the redshift bin.

\subsection{The Fisher information with foreground-wedge avoidance}
\label{sec:fisher_wedge}

When the foreground-wedge avoidance is applied, the usual approach based on the isotropic integral over the Fourier modes of $k$ for power spectrum and over triangles $(k_1,k_2,k_3)$ for (reduced) bispectrum are no longer adequate.  The wedge introduces an anisotropic mask that removes the $k$-modes located in $k_{\parallel}\leq m(z)\, k_{\perp}$, thereby altering the number of available modes in Fourier space. To accurately incorporate the effects of the wedge, we adopt a more general Fisher forecast formalism that explicitly sums over the perpendicular and parallel components of $(k_{1\perp},k_{2\perp},k_{3\perp},k_{1\parallel},k_{2\parallel})$, which  accommodates the anisotropic mode removal. We don't use $k_{3\parallel}$ as the sixth dimension because it is determined by the relation  
$k_{3\parallel}=-k_{1\parallel}-k_{2\parallel}$. To satisfy 
the intrinsic relation $ \boldsymbol k_1 + \boldsymbol k_2 + \boldsymbol k_3= \boldsymbol 0$, we have $ \boldsymbol k_{1\perp}+ \boldsymbol k_{2\perp}+ \boldsymbol k_{3\perp}= \boldsymbol 0$ and $k_{1\parallel}+k_{2\parallel}+k_{3\parallel}=0 $, where $k_{i\perp}=|\boldsymbol k_{i\perp}|$ ($i=1,2,3$). In our convention, $k_{\parallel,\rm min}$, $k_{\parallel,\rm max}$, $k_{\perp,\rm min}$, and $k_{\perp,\rm max}$ are all positive constants, and $k_{i \perp}>0$, while $k_{i\parallel}$ could be either positive or negative representing parallel or anti-parallel to the line-of-sight direction. Additionally, it is necessary to 
 modify the Fourier space volume and the number of independent modes
in a way that properly accounts for the exclusion of wedge-contaminated $k$-modes in Fourier space, ensuring unbiased and precise Fisher forecasts.

We adopt the \textit{integral form} of the Fisher matrix rather than an explicit mode--by--mode summation. This choice reduces the computational cost, as the integration  can be evaluated efficiently. We have checked the numerical convergence of these integrals to ensure stable results. We numerically evaluate  the Fisher matrix 
and impose the foreground-wedge mask directly on the numerical integration to exclude the wedge-contaminated region in Fourier space. 
For the power spectrum, we have
\begin{equation}
F_{\alpha\beta}^{P}(z_i)=\frac{1}{2}
\int_{k_{\perp,\min}}^{k_{\perp,\max}}\!\!dk_{\perp}
\left(\int_{-k_{\parallel,\max}}^{-k_{\parallel,\min}}\!+\int_{k_{\parallel,\min}}^{k_{\parallel,\max}}\!\right)\ dk_{\parallel}
\ \frac{ \mathcal F_\mathrm{P}(k_{\perp},
k_{\parallel},z_i)}{\mathrm{d}k_\perp \mathrm{d}k_\parallel},
\label{eq:fisher_2d_ps}
\end{equation}
\begin{equation}
\mathcal F_\mathrm{P}(k_{\perp},k_{\parallel},z_i)=
\frac{\partial_\alpha  P_{21} \partial_\beta P_{21}}{\Delta P_{\rm tot}^2}=\frac{\partial_\alpha  P_{21}\partial_\beta P_{21}}{P_{\rm tot}^2/N_P},
\end{equation}
where $N_{\rm P}$ is the number of independent modes in that Fourier bin, i.e.  
$$N_{\rm P} = (V_{\rm survey}/V_{\rm one})\times (\frac{2\pi k_{\perp}\mathrm{d}k_\perp \mathrm{d}k_\parallel  }{k_{\perp,\mathrm{min}}^2 k_{\parallel,\mathrm{min}}})=\frac{V_{\rm survey}\,k_{\perp}\mathrm{d}k_\perp \mathrm{d}k_\parallel }{(2\pi)^2},$$
and 
$$V_{\rm one}=\frac{(2\pi)^3}{k_{\perp,\mathrm{min}}^2 k_{\parallel,\mathrm{min}}}.$$ 
In the first equation calculating $N_{\rm P}$,
the first factor $(V_{\rm survey}/V_{\rm one})$ corresponds to the number of independent observations, and the second factor counts the number of independent modes in a thin cylindrical shell with volume of $2\pi k_\perp \mathrm{d}k_\perp \mathrm{d}k_\parallel$
determined by $(k_\perp , k_\parallel)$. We divide the shell by the fundamental cell$\ k_{\perp,\mathrm{min}}^2 k_{\parallel,\mathrm{min}}\ $, and we also add a factor $1/2$ before the total integration to account for redundant $k$-modes between $\pm \vec k$.

For the reduced bispectrum, we have
\begin{equation}
\begin{split}
F_{\alpha\beta}^{Q}(z_i)=&
\frac{1}{6}~
\!\!\int_{k_{\perp,\min}}^{k_{\perp,\max}}\!\!dk_{1\perp}
\int_{k_{\perp,\min}}^{k_{\perp,\max}}\!\!dk_{2\perp}
\int_{k_{\perp,\min}}^{k_{\perp,\max}}\!\!dk_{3\perp}
\\&\times \left(
\int_{-k_{\parallel,\max}}^{-k_{\parallel,\min}}\!+
\int_{k_{\parallel,\min}}^{k_{\parallel,\max}}\!\right)dk_{1\parallel}
\left(
\int_{-k_{\parallel,\max}}^{-k_{\parallel,\min}}\!+
\int_{k_{\parallel,\min}}^{k_{\parallel,\max}}\!\right)dk_{2\parallel}\ \frac{\mathcal F(k_{1\perp},k_{2\perp},k_{3\perp},
k_{1\parallel},k_{2\parallel},z_i)}{dk_{1\perp}dk_{2\perp}dk_{3\perp}dk_{1\parallel}dk_{2\parallel}},
\label{eq:fisher_5d_kpkpkl}
\end{split}
\end{equation}
with 
\begin{equation}
\mathcal F=\frac{\partial_{\alpha} Q_{21}\ \partial_{\beta} Q_{21}}{\frac{(2\pi)^3 }{(k_{\perp, \rm min}^2 k_{\parallel, \rm min})N_Q}P_{\rm tot}(k_{1\perp},k_{1\parallel},z_{i})P_{\rm tot}(k_{2\perp},k_{2\parallel},z_{i})P_{\rm tot}(k_{3\perp},k_{3\parallel},z_{i})/\left[ P _ { 21 } \left( k _ { 1 } \right) P _ { 21 } \left( k _ { 2 } \right) + ( 2\, perm. ) \right]^2}
\end{equation}
Note the factor 1/6 accounts for the redundant $k$-modes \citep{Bull2024}, so there is no factor $s_{123}=6,2,1$ in the variance of reduced bispectrum.

The definition of $N_Q$ for the reduced bispectrum is similarly modified, 
\begin{equation}
    N_Q=(\frac{V_\mathrm{survey}}{V_{\rm one}})\times (\frac{V_Q }{k_{\perp,\mathrm{min}}^4 k_{\parallel,\mathrm{min}}^2})=\frac{V_\mathrm{survey}V_Q}{(2\pi)^3({k_{\perp,\mathrm{min}}^2 k_{\parallel,\mathrm{min}})}}
\label{N_Q}
\end{equation}
where $V_Q$ is the thin shell volume for fixed $(k_{1\perp},k_{2\perp},k_{3\perp},k_{1\parallel},k_{2\parallel})$ of the reduced bispectrum. The derivation of $V_Q$ is given in the appendix, and we have
\begin{equation}
V_{Q}= \frac{2\pi k_{1\perp}k_{2\perp}k_{3\perp}dk_{1\perp}dk_{2\perp}dk_{3\perp}dk_{1\parallel}dk_{2\parallel}}{\Delta_k}, 
\label{V_Q}
\end{equation}
where 
\begin{equation}
\Delta_k=\frac{1}{4}\sqrt{\left[k_{3\perp}^2-(k_{1\perp}-k_{2\perp})^2\right]\left[(k_{1\perp}+k_{2\perp})^2-k_{3\perp}^2\right]},
\end{equation}
and $k_{3\parallel}=-k_{1\parallel}-k_{2\parallel}$.

\subsection{The interferometer array setup}
\label{sec:intrumentations}
To ease comparison with previous results, 
we adopt the same redshift range of $22\le z\le 182$ as in \cite{Bull2024},  and divide it into ten bins of equal radial comoving distance. However, we will show below that 
the higher redshift bins have negligible contribution to the Fisher information, and the main constraining power is from lower redshift bins. Assuming a spatially flat universe, transverse Fourier modes $k_{\perp}$ are related to interferometric baselines via
\begin{equation}
k_\perp = \frac{2\pi |\boldsymbol{u}|}{\chi(z)}, \qquad {\rm with} \  \boldsymbol{u}\equiv \frac{\boldsymbol{D}}{\lambda_{\rm obs}}, 
\end{equation}
where $\chi(z)$ is the radial comoving distance, $\boldsymbol{D}$ is the physical baseline vector, and $\lambda_{\rm obs}$ is the observing wavelength of the redshifted 21-cm line, i.e. $ \lambda_{\rm obs}=\lambda_{21}(1+z)$.

We adopt the approximate baseline distribution of a lunar array following \cite{Bull2024}, which models the 1-D baseline density as
\begin{equation}
n_d(D) = \mathcal{A}\,(D-D_0)^2 \exp\!\left[-\left(\frac{D-D_0}{w}\right)^2\right],
\end{equation}
where the $w$ and $D_0$ are free parameters and $\mathcal{A}$ is a normalization constant determined by
\begin{equation}
\int_{D_{\rm min}}^{D_{\rm max}} 2\pi D\, n_d(D)\,\mathrm{d}D \;=\; \frac{N_{\rm ant}(N_{\rm ant}-1)}{2},
\end{equation}
with $N_{\rm ant}$ the total number of antennas. 
The two-dimensional $uv$-plane density $n_b(\boldsymbol{u})$ is related to $n_d$ through
\begin{equation}
\int n_b(\boldsymbol{u},z)\,\mathrm{d}\boldsymbol{u}
= \int 2\pi D\, n_d(D)\,\mathrm{d}D,
\qquad \mathrm{where}\  D(\boldsymbol u,z)=\lambda_{\rm obs}|\boldsymbol{u}|=\lambda_{21}(1+z)|\boldsymbol{u}|,
\end{equation}
then we can calculate $n_b(\boldsymbol u, z)$ as 
\begin{equation}
    n_b(\boldsymbol u,z)= \lambda_{21}^2 (1+z)^2 n_d (D).
\end{equation}

Therefore, the thermal-noise contribution to the 21-cm brightness-temperature power spectrum is written as \citep{Bull2024}
\begin{equation}
P_N(k_{\perp},z) =
T_{\rm sys}^2\,
\frac{\lambda_{\rm obs}^5\,\chi^{2}(z)\,(1+z)\Omega_{\rm area}\,}
{A_{\rm eff}^2\,H(z)\,t_{\rm tot}\,n_b(\boldsymbol{u},z)\,N_{\rm pol}\,\Omega_{\rm FOV}},
\end{equation}
where $T_{\rm sys}=T_{\rm sky} + T_{\rm inst}\approx T_{\rm sky}\approx5000~\mathrm{K}\,(\nu/50~\mathrm{MHz})^{-2.5}$ is the brightness-temperature of the Milky Way foreground, 
$A_{\rm eff}=2\lambda_{\rm obs}^2/(4\pi)$ is the effective collecting area for each dipole antenna, $H(z)$ is the Hubble parameter, $N_{\rm pol}=2$ is the number of measured polarizations, $\Omega_{\rm FOV}=\pi$ is the field of view of each dipole antenna, and $\Omega_{\rm area}$ is the solid angle of the total survey area.   
We assume a uniform coverage of the sky, so that the effective integration time per  observation $V_{\rm one}$ is $t_{\rm pix}=t_{\rm tot}\,\Omega_{\rm FOV}/\Omega_{\rm area}$. 

\begin{figure*}[htbp]
  \centering
  \includegraphics[width=\textwidth]{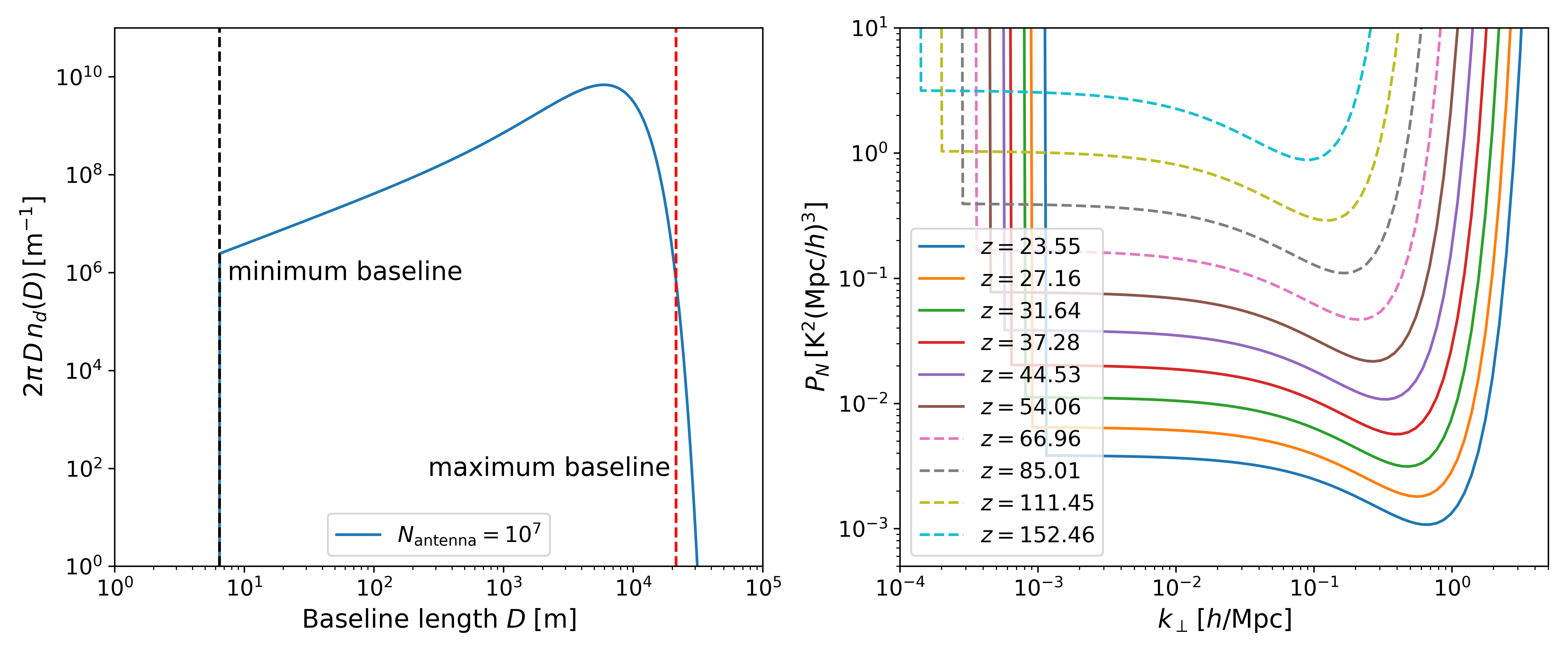}
  \caption{
\emph{Left:} Baseline density distribution
$2\pi D\,n_{\rm d}(D)$ for an array with $N_{\mathrm{antenna}}=10^{7}$ elements. 
  The black dashed line gives the shortest baseline set by the half-wavelength 
  ($\lambda_{\rm obs}(z=60)/2$), and the red dashed line indicates the
  diameter of a hypothetical close‑packed hexagonal array, i.e.\ the longest usable
  baseline.  
\emph{Right:} Thermal‑noise power spectrum
$P_{N}(k_\perp)$ evaluated at the \textit{central redshift}
of each redshift bin. 
All bins have the same radial comoving width. The six solid curves are the noise spectrum for
the bins effectively contributed in our Fisher analysis, whereas the dashed curves
illustrate the noise level in higher‑$z$ bins that are essentially negligible
in the forecast.  
  }
\label{fig:baseline_noise}
\end{figure*}

Due to the very high sky temperature at the low frequencies, an extremely large interferometer array is required to obtain high signal to noise ratio in this measurement. For our fiducial configuration, we take $D_0=-2100~\mathrm{m}$, $w=6200~\mathrm{m}$, and $N_{\rm ant}=10^7$. Our fiducial survey covers $20{,}000~\mathrm{deg}^2$ with a total integration time of 22000 hours as in \cite{Bull2024}.

Fig.~\ref{fig:baseline_noise} shows the baseline density and noise power spectrum of the hypothetical array. The minimum baseline is set to be the half-wavelength at redshift $z=60$, which sets the lowest transverse mode we can probe, while the close-packed diameter defines the high-$k_\perp$ cutoff.  Between these limits the baseline
distribution peaks near $d\!\sim\!10^{4}$ m, leading to the smallest noise at $k_\perp\!\sim\!0.1-1\;h\,\mathrm{Mpc^{-1}}$ for the six bins (solid curves) that enter our Fisher forecast. 

At the higher central redshifts, which are shown as dashed curves in the right panel of Fig.~\ref{fig:baseline_noise}, the sky temperature is too high, so the noise levels are elevated to very high value. Therefore, the contribution of these high redshift bins to the cosmological constraint are negligible.

\section{Results}\label{sec:results}

We now use the anisotropic Fisher matrix formalism with the foreground wedge avoidance,
described in the previous section, 
to forecast the sensitivity of the 21-cm power spectrum and reduced bispectrum to the primordial oscillatory features. In the cylindrical Fourier-space formalism, we compute the total signal-to-noise ratio (SNR) and construct Fisher forecasts for both observables. 

\begin{figure}[htbp]
  \centering
  \includegraphics[width=0.6\textwidth]{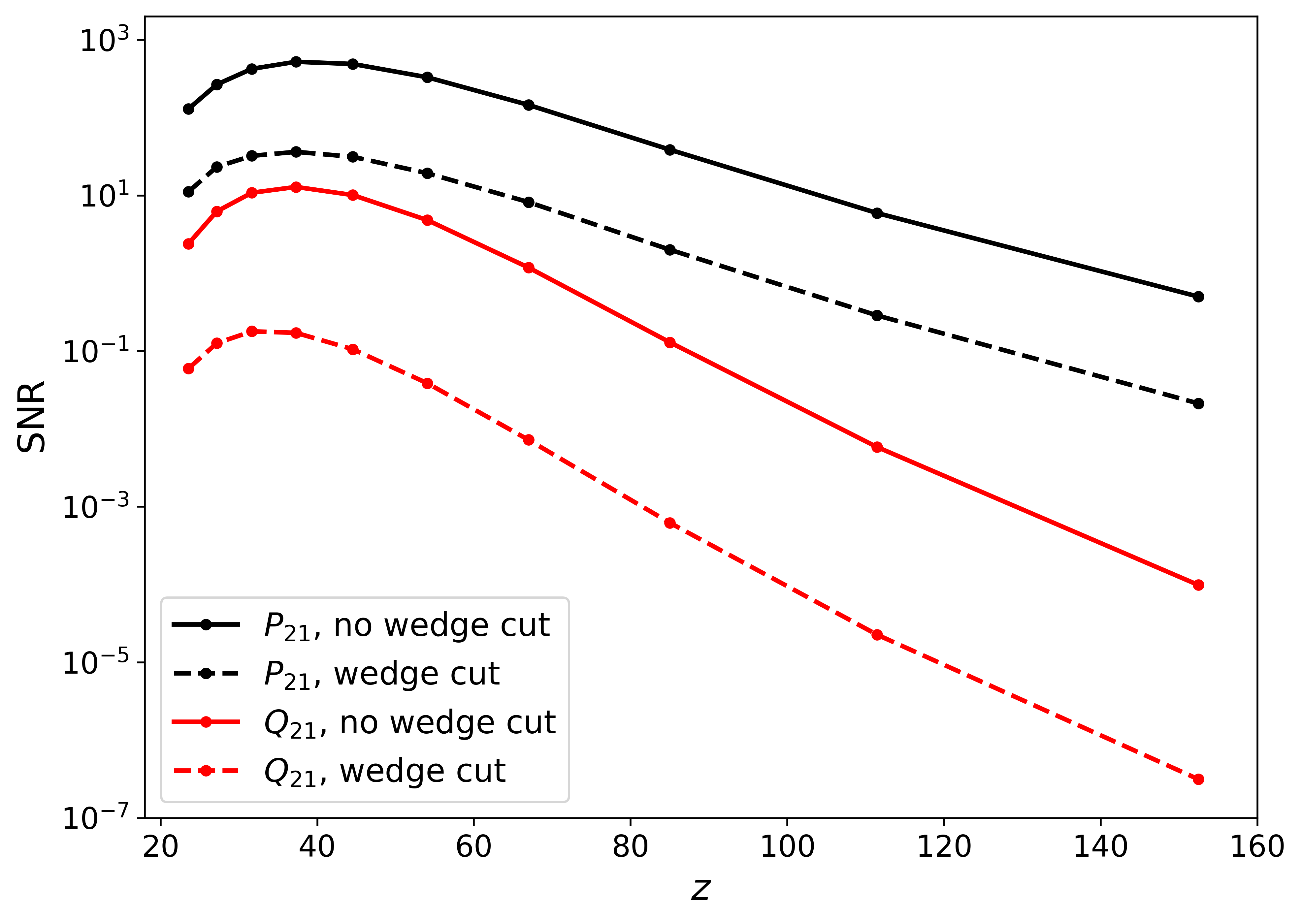}
  \caption{
Signal-to-noise ratio (SNR) of the  
21-cm power spectrum $P_{21}$ measurements (black curves) and the 
reduced 21-cm bispectrum $Q_{21}$ measurements (red curves) in the standard model, evaluated at the central redshift of each $z$-bin. 
The solid lines are the results \emph{without} foreground-wedge excision (no wedge cut), while the dashed lines correspond to results \emph{with} foreground-wedge excision (wedge cut).
    }
  \label{fig:snr_profiles}
\end{figure}

Following \cite{Bull2024}, we divide the redshift range $22 \leq z \leq 182$ into ten bins with equal comoving-distance interval. 
For each $z$-bin, we sum over the square of SNR of all available $k$-bins, and take the square root of the total SNR$^2$ to approximate the total SNR$_i$ for $i$-th $z$-bin.
The resulting SNR obtained for the 21-cm power spectrum and the reduced 21-cm bispectrum are shown in Fig.~\ref{fig:snr_profiles}, for both cases of with and without the foreground wedge excision.
In each case, the SNR first increases with redshift due to a larger comoving volume that could be
probed, reaches a maximum at $z \simeq 35-40$, and then decreases rapidly toward higher redshifts mainly due to increasingly higher foreground temperature and hence higher thermal noise at lower frequencies. The power spectrum gives substantially higher SNR than the reduced bispectrum over the full redshift range considered. The decline is particularly steep for the reduced bispectrum at high redshift, especially when the foreground wedge is excluded.  

At a fixed redshift, the dashed curves are always below the corresponding solid ones, showing that foreground-wedge excision reduces the total SNR significantly. This reduction is expected, since excluding the wedge modes reduces the number of accessible Fourier modes. We take the wedge-cut case as the conservative limit; if the foreground removal technique is effective, some of the 21-cm signal in the foreground wedge can be recovered \citep{Gagnon2021,Li2024}, so the achievable sensitivity is expected to lie between the wedge-cut and no-wedge-cut cases.

Since the redshift bins are independent of each other, we obtain the total SNR by summing over their contributions to $\mathrm{SNR}^2$,
\begin{equation}
\mathrm{SNR}_{\mathrm{tot}}^{2} = \sum_{i=1}^{N_{\mathrm{bin}}} \mathrm{SNR}_{i}^{2}.
\label{eq:snr_tot}
\end{equation}
We find that the six bins at lower redshift ($22\le z\le 60$) contribute $>\!90\%$ of the total SNR for all four cases and $>\!95\%$ for the reduces 21-cm bispectrum. Accordingly, the following Fisher forecasts focus on these six bins, which contain most of the constraining power.

\begin{figure}[htbp]
  \centering
  \includegraphics[width=0.6\linewidth]{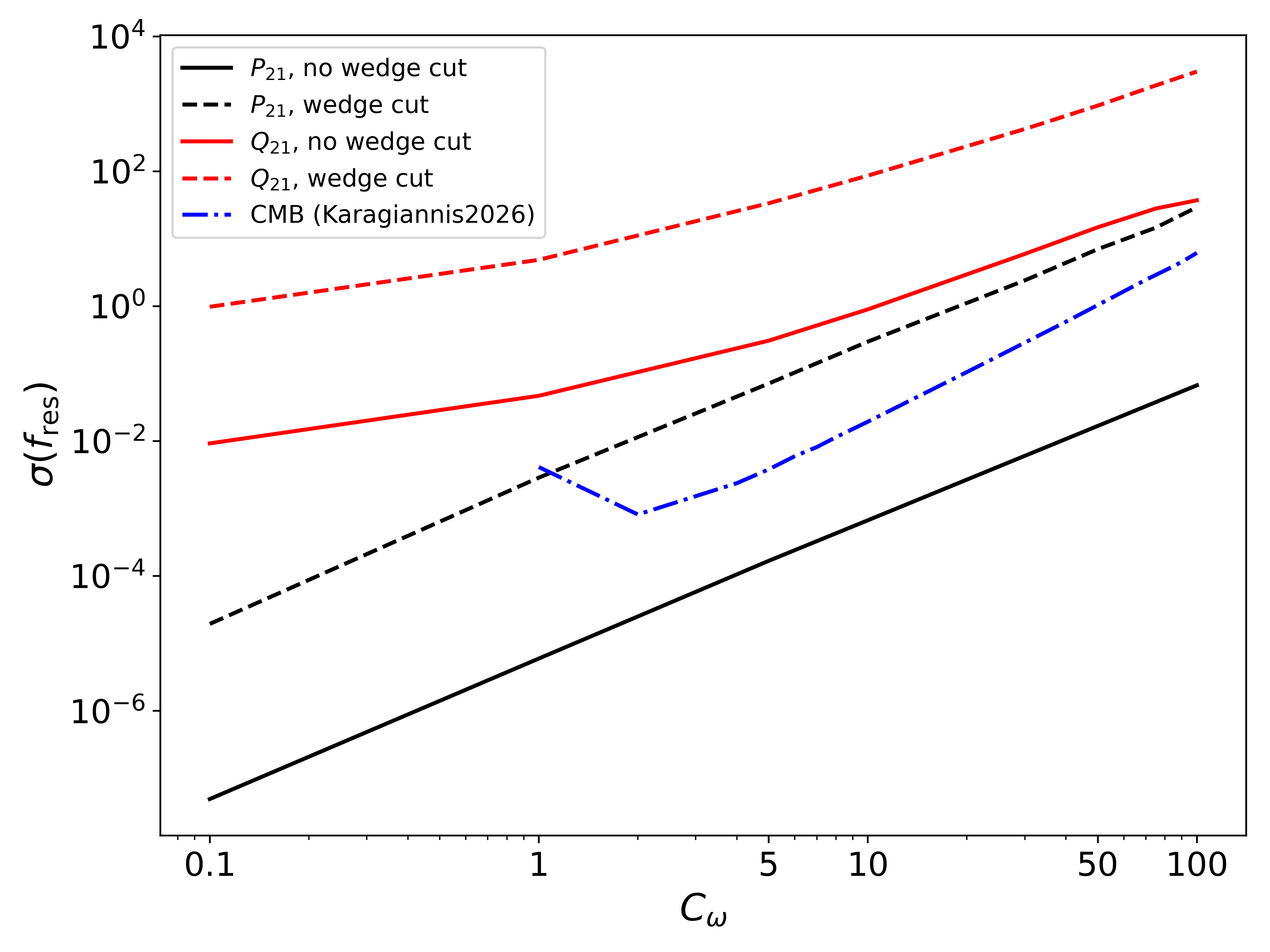}
  \caption{
1-$\sigma$ error expected for the primordial oscillation  
amplitude $f_{\mathrm{res}}$ as a function of the oscillation frequency
$C_{\omega}$ in the resonant model, using 21-cm power spectrum measurements (black lines) and 21-cm bispectrum measurements (red lines) from the Dark Ages, respectively.
The solid (dashed) line corresponds to the result retaining (removing) the modes in the foreground‑wedge.
For comparison, we also plot the forecasted constraints from CMB observations combining SO and LiteBIRD with the blue dot-dashed line, using data adapted from \cite{Karagiannis2026}.
}
  \label{fig:sigma_Cw}
\end{figure}

We then apply the Fisher analysis to forecast constraints on the oscillatory 
amplitude parameters in the resonant and step feature models, respectively, assuming fiducial amplitudes $f_{\rm res}=0$ and $\epsilon_{\rm step}=0$, and quantifies the impact of foreground-wedge avoidance strategy.
Fig.~\ref{fig:sigma_Cw} shows the \(1\sigma\) uncertainty expected for the resonant-model 
parameter \(f_{\mathrm{res}}\) as a function of the oscillation frequency parameter \(C_{\omega}\). As \(C_{\omega}\) increases, corresponding to a transition from slowly to rapidly oscillating 
features, the precision of \(f_{\mathrm{res}}\) steadily degrades for both the power spectrum 
($P_{21}$) and the reduced bispectrum 
($Q_{21}$) observables. 
Note that a larger $C_{\omega}$ gives more oscillations within the available $k$-range. However, the effective amplitude of the resonant feature scales as $A_{\log} = 8 f_{\mathrm{res}}/C_{\omega}^2$, as expected from Eq.~(\ref{PP_res}), so a larger $C_{\omega}$ also means a more suppressed amplitude of the oscillatory signal and hence a lower effective SNR. 
The scaling of the uncertainty expected from $P_{21}$ measurements shown in Fig.~\ref{fig:sigma_Cw} implies that \(\sigma(A_{\log})\) remains nearly constant across the entire \(C_{\omega}\) range explored.

In all cases, $P_{21}$ provides a tighter constraint on \(f_{\mathrm{res}}\) compared to $Q_{21}$, primarily because the SNR for $P_{21}$ is 2 to 3 orders of magnitude higher than that for $Q_{21}$, as seen in Fig.~\ref{fig:snr_profiles}.
Removing the foreground wedge modes reduces the number of available \(k\)-modes significantly, leading to an increase in the uncertainties by approximately two orders of magnitude for both $P_{21}$ and $Q_{21}$ observables. 
The dot-dashed line  in Fig.~\ref{fig:sigma_Cw} shows the expected uncertainties of \(f_{\mathrm{res}}\) from the next-generation CMB observations with the Simons Observatory (SO) and LiteBIRD, including temperature anisotropies, E-mode polarisation, and the CMB lensing observables (data taken from \cite{Karagiannis2026}).
It shows that with the 3D information retained in 21-cm intensity mapping observations, and the large number of $k-$modes available from the Dark Ages, the 21-cm power spectrum measurements can potentially achieve a higher precision for $f_{\rm res}$ than the CMB observations if the foreground can be effectively removed. However, if the modes in the foreground wedge cannot be recovered, then the foreground avoidance strategy would significantly degenerate the constraining power, making the Dark Ages 21-cm observations less promising. 

\begin{figure}[htbp]
  \centering
  \includegraphics[width=1.0\linewidth]{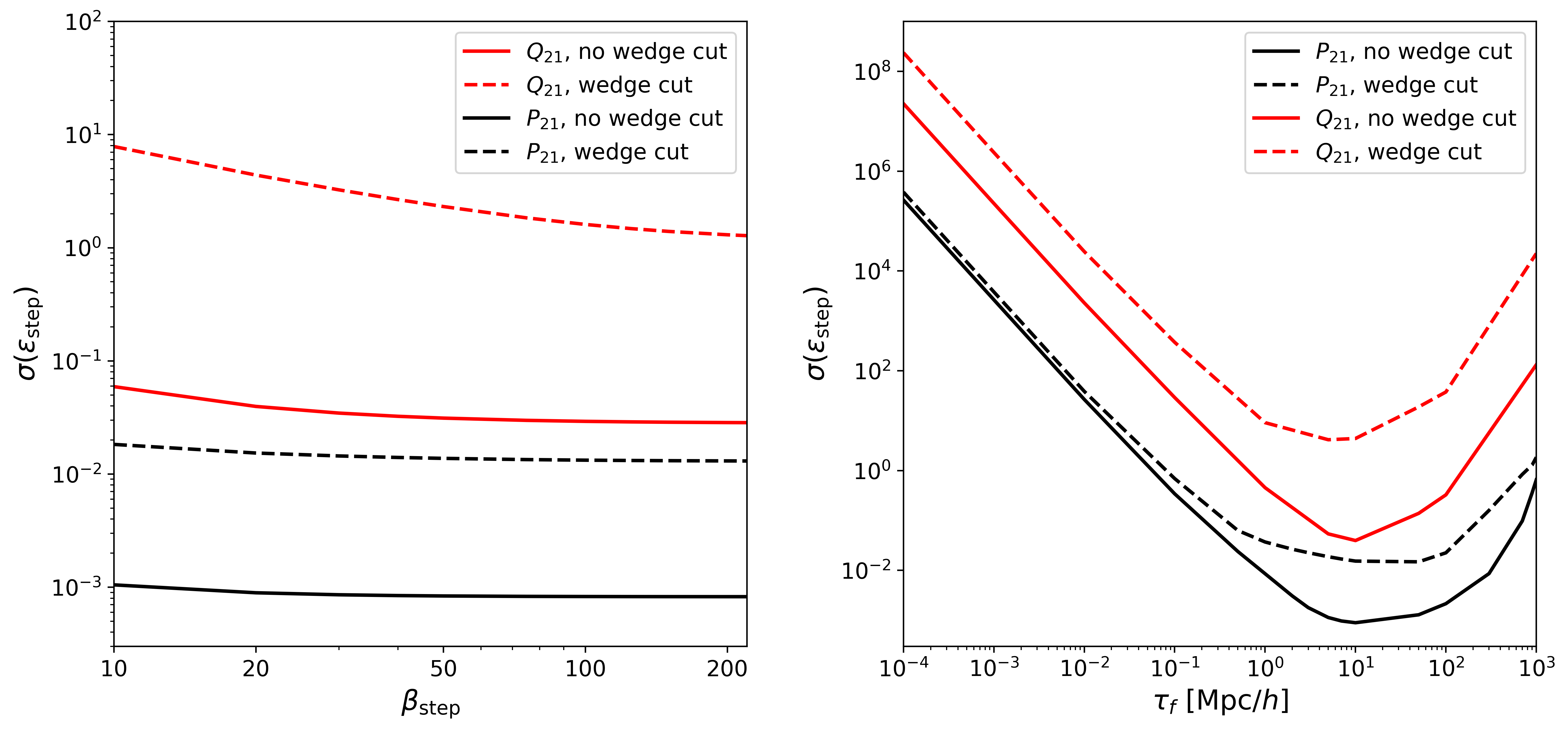}
  \caption{
1-$\sigma$ error expected for the height of step in the inflaton potential
$\epsilon_{\rm step}$
for the step model, assuming a fiducial $\epsilon_{\rm step}=0$.
\textit{Left:} $\sigma_{\epsilon_{\rm step}}$ as a function of $\beta_{\rm step}$ at fixed $\tau_f=10\,{\rm Mpc}\,h^{-1}$. 
\textit{Right:} $\sigma_{\epsilon_{\rm step}}$ as a function of $\tau_f$
at fixed $\beta_{\rm step}=20$. 
In each panel, the black lines are the predicted precision achievable using 21-cm power spectrum measurements from the Dark Ages, and the red lines are the precision expected for 21-cm bispectrum measurements.
The solid (dashed) line corresponds to the result retaining
(removing) the modes in the foreground‑wedge.
  }
\label{fig:sigma_betas_tauf}
\end{figure}

Fig.~\ref{fig:sigma_betas_tauf} shows the constraint on the step-feature model amplitude parameter $\epsilon_{\rm step}$ for different $\beta_{\rm step}$ and $\tau_f$ values, respectively. The constraint on $\epsilon_{\rm step}$ tightens slightly with increasing $\beta_{\rm step}$, but varies non-monotonically with $\tau_f$, reaching its strongest value around $\tau_f\sim10\ {\rm Mpc/h}$. 
This sensitive scale is consistent with the scale that our hypothetical array is most sensitive to (Fig.~\ref{fig:baseline_noise}). 
The array for the Dark Ages would yield weaker constraints on the feature amplitude for very small or very large $\tau_f$.

In the step model, when the wedge is removed, uncertainties also increase for both estimators across the scanned parameter ranges. However, we find that there are some regions in the parameter space where wedge excision has little impact on the 21-cm power spectrum signal (the small-scale part of black curves in the right panel of Fig.~\ref{fig:sigma_betas_tauf}). The oscillatory $P_{21}$ signal is concentrated near a characteristic scale set by \(\tau_f\) (roughly \(k\sim \tau_f^{-1}\)), in contrast to the resonant case where the resonant modulation extends over the entire range of observable scales.
When this step feature lies largely outside the foreground-affected region of \((k_\perp,k_\parallel)\) space, removing the wedge would then have a much weaker impact on the parameter constraints. 
Consequently, the \(1\sigma\) uncertainties in $\epsilon_{\rm step}$ are nearly the same for \(\tau_f \lesssim 0.1\, {\rm Mpc/h}\) in $P_{21}$ measurements with or without the wedge cut.

\begin{figure}[htbp]
  \centering
\includegraphics[width=1.0\linewidth]{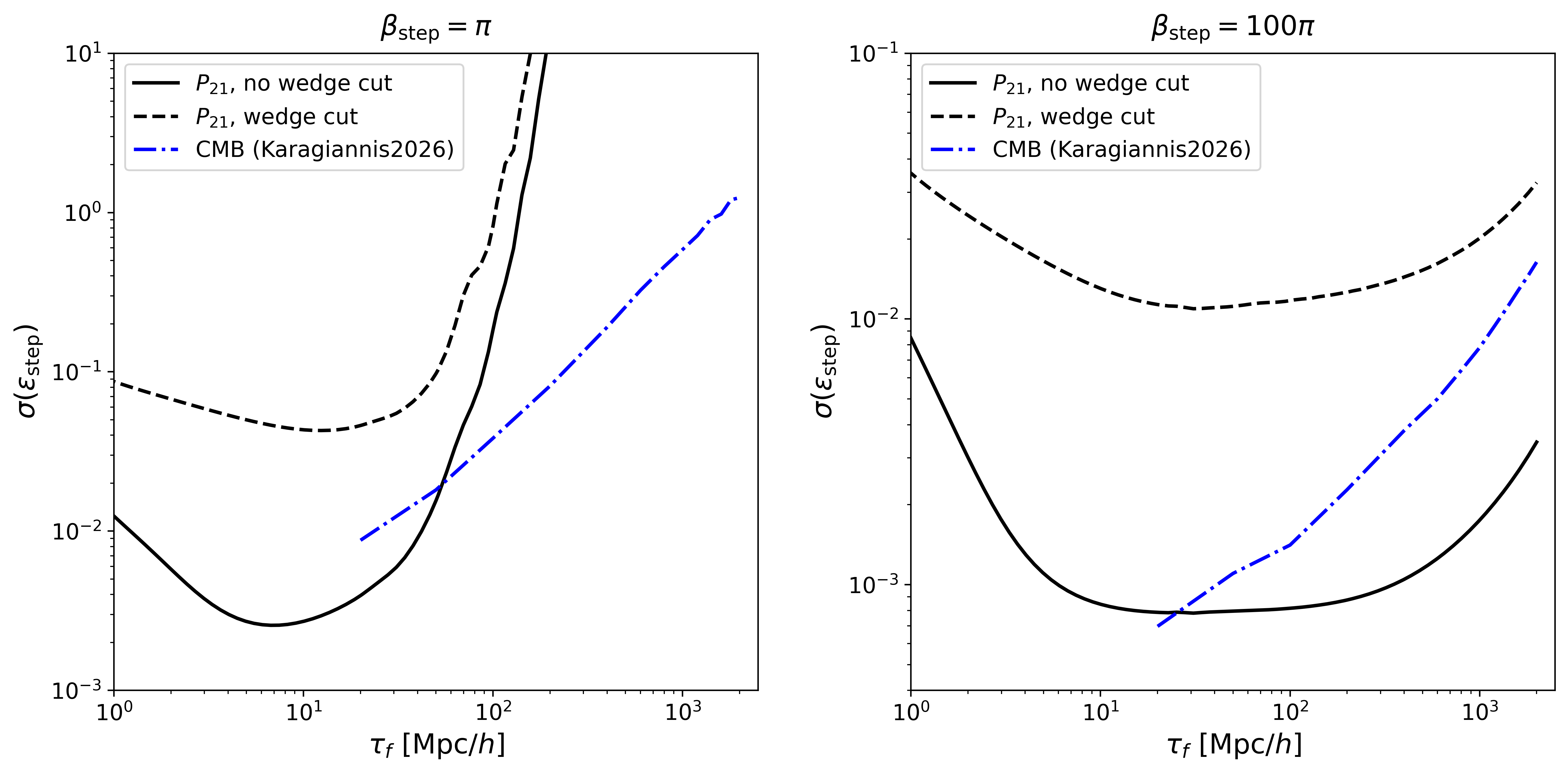}
  \caption{
    1-$\sigma$ error expected for the height of step in the inflaton potential $\epsilon_{\rm step}$ for the step model, assuming a fiducial $\epsilon_{\rm step}=0$.
    \textit{Left:} $\sigma_{\epsilon_{\rm step}}$ as a function of $\tau_f$ at fixed $\beta_{\rm step}=\pi$. 
    \textit{Right:} $\sigma_{\epsilon_{\rm step}}$ as a function of $\tau_f$
    at fixed $\beta_{\rm step}=100\pi$. 
    In each panel, the black curves show the forecasted constraints from the Dark-Ages 21-cm power spectrum measurements, with the solid (dashed) curve retaining (removing) modes within the foreground wedge.
    For comparison, we also plot the forecasted constraints from CMB observations combining SO and LiteBIRD with the blue dot-dashed line in each panel, using data adapted from \cite{Karagiannis2026}.
    }
\label{fig:step_CMB}
\end{figure}

Currently, no statistically robust evidence for inflationary feature models beyond a nearly power-law primordial spectrum is found, either in the latest Planck data of the CMB \citep{Planck2018} or in galaxy surveys \citep{Mergulhao2023}.
To compare our results with the next generation CMB constraints, we plot the $1\sigma$ forecast errors on $\epsilon_{\rm step}$ as a function of $\tau_f$ for $\beta_{\rm step}=\pi$ and $\beta_{\rm step}=100\pi$ respectively in Fig.~\ref{fig:step_CMB}, taking the same sharpness parameters as in \cite{Karagiannis2026}. The corresponding CMB constraints expected from combining SO and LiteBIRD are plotted with dot-dashed curves for comparison (data adapted from \cite{Karagiannis2026}).
If the step is not sharp ($\beta_{\rm step}$ is small, as in the left panel), the CMB is more sensitive on large scales, resulting in higher precision on $\epsilon_{\rm step}$ for $\tau_f\gtrsim 60\, {\rm Mpc/h}$. For a steep step (large $\beta_{\rm step}$, as in the right panel), or $\tau_f \lesssim 60\, {\rm Mpc/h}$ in the small $\beta_{\rm step}$ case, the measurement error depends on the foreground removal capability. The 21-cm power spectrum can be much more sensitive to the step feature if foreground can be effectively removed, while in the case that the foreground-contaminated modes have to be fully avoided, the sensitivity would be lower than the next generation CMB experiments. 
In any case, the 21-cm observations from the Dark Ages can probe smaller scales than the CMB, providing useful information on complementary parameter space.

\section{Conclusion and Discussions}
\label{sec:conclusions}

In this work, we use a Fisher-matrix framework to forecast the constraints of inflationary features  
from 21-cm intensity mapping during the Dark Ages, with a focus on inflation models with oscillatory features, for which the three-dimensional information encoded in intensity mapping observations may potentially improve the constraints from the two-dimensional CMB constraints. 
In particular, we improve the traditional Fisher-matrix formalism to self-consistently account for the impact of foreground wedge avoidance on both the cylindrical 21-cm power spectrum ($P_{21}$) and the reduced 21-cm bispectrum ($Q_{21}$) measurements, and also take into account
the impact of thermal noise that is extremely large for Dark Ages 21-cm experiments. This treatment gives a more reasonable estimate of the achievable precision by propagating the loss of foreground-wedge modes to the loss of sensitivity on specific scales. 

We find that excluding the foreground wedge leads to a significant reduction in the range of available $k$-modes, especially in the Dark Ages when this effect is much more severe than in lower-redshift surveys. 
It is found that the foreground avoidance strategy reduces the total signal-to-noise ratio by roughly two orders of magnitude for the redshift range considered here. 
We note that the foreground-avoidance strategy has a
stronger impact on the parameter sensitivity in our analysis than predicted in \cite{Mondal2023}. This is mainly due to the larger array scale and longer available baselines adopted in this paper. Compared with the instrumental configuration in \cite{Mondal2023}, our longer baselines extend the measurement to larger $k_\perp$, where the foreground-wedge avoidance removes a larger fraction of the accessible $k$-modes, so our prediction is more consistent with the pessimistic case in \cite{Mondal2023}.

In our analysis,  the 21-cm power spectrum provides much stronger constraints on the feature amplitude 
parameters than the reduced bispectrum, primarily due to the larger signal amplitude in the power spectrum. 
It implies that for both the resonant model and the step model, primordial non-Gaussianity is not significant in constraining the models.
Nevertheless, the reduced bispectrum remains useful for signals with significant three-point correlations, and can provide an independent cross-check on power-spectrum results. The wedge-aware treatment developed here can therefore be applied to other 21-cm bispectrum measurements, including other high-redshift 21-cm experiments for the cosmic dawn and epoch of reionization.

The 1-$\sigma$ measurement errors on the feature amplitude parameters presented here should be considered as  conservative estimates, for the nominal array size and survey parameters as adopted in \cite{Bull2024}. In principle, the signal-to-noise ratio of the power spectrum or the bispectrum measurements can be further improved by   
distributing the total collecting area into multiple stations tuning the array configuration to optimize for a range of scales of interest \citep{WenYuewei2026}.
Based on our forecasted measurement precision of feature amplitude parameters, 
21-cm power spectrum measurements from the Dark Ages could improve on the next-generation CMB constraints by about an order of magnitude for the resonant model, provided that the foreground can be removed effectively.
In the step model, however, the constraining power, as compared with the CMB, depends on the model parameters. 
A smaller \(\tau_f\) shifts the feature to smaller scales, and a larger \(\beta_{\rm step}\) makes the high-\(k\) oscillations less damped, both correspond to a parameter space to be better constrained by the 21-cm observations from the Dark Ages, but the exact constraints will be subject to foreground removal capability.
For both categories of models, the 21-cm intensity mapping observations can probe smaller scales than the CMB, and have the potential to complement the CMB observation by covering a different parameter space. 
These results highlight the potential of 21-cm intensity mapping as a probe of the early Universe and also the importance of careful foreground removal in future forecasts and real observations.

\begin{acknowledgments}
We thank Meng Zhou and Xingang Chen for helpful discussions. This work was supported by the NSFC grant No. 12533001, the National Key R\&D Program of China No. 2022YFF0504300, China's Space Origins Exploration Program Nos. GJ11010405 and GJ11010401, the NSFC International (Regional) Cooperation and Exchange Project No. 12361141814, and by the Specialized Research Fund for State Key Laboratory of Radio Astronomy and Technology.
\end{acknowledgments}

{\Large {\it Softwares} used:} {Colossus \citep{2018ApJS..239...35D},
Astropy \citep{2013A&A...558A..33A,2018AJ....156..123A,2022ApJ...935..167A},
NumPy \citep{2020Natur.585..357H},
CAMB \citep{2000ApJ...538..473L}.}


\appendix
\section{The Fourier volume factor $V_Q$}

The factor $V_Q$ in Eq.~\eqref{V_Q} denotes the Fourier-space configuration volume of a finite cylindrical triangle bin. After division by the appropriate fundamental Fourier-space cell volume, it gives $N_Q$ in Eq.~\eqref{N_Q}, which is the number of fundamental triangle configurations in the bin. The factor $V_Q$ can be derived as follows.  

We write each Fourier mode as
\begin{equation}
    \boldsymbol{q}_i =
    \boldsymbol{q}_{i\perp} +q_{i\parallel}\hat{\boldsymbol{n}},
    \qquad
    q_{i\perp}\equiv |\boldsymbol{q}_{i\perp}|>0 \quad (i=1,2,3)
\end{equation}
where $\hat{\boldsymbol{n}}$ is fixed unit vector along the line-of-sight direction, $q_{i\parallel}$ is the line-of-sight component which could be either positive or negative, and $q_{i\perp}$ is the $k$-mode perpendicular to line-of-sight. The cylindrical triangle bin is specified by
\begin{equation}
    q_{i\parallel}\in 
    \left[
    k_{i\parallel}-\frac{\Delta k_{i\parallel}}{2},
    k_{i\parallel}+\frac{\Delta k_{i\parallel}}{2}
    \right]\ (i=1,2),
    \qquad
    q_{i\perp}\in 
    \left[
    k_{i\perp}-\frac{\Delta k_{i\perp}}{2},
    k_{i\perp}+\frac{\Delta k_{i\perp}}{2}
    \right]\ (i=1,2,3).
\end{equation}
The triangle condition $\boldsymbol{q}_1+\boldsymbol{q}_2+\boldsymbol{q}_3=\boldsymbol 0$ can be imposed by the three-dimensional Dirac delta function $\delta_D^{(3)}$, and
therefore, we can write
\begin{equation}
\begin{aligned}
    V_Q
    &=
    \int_{\rm bin} \mathrm{d}^3\boldsymbol{q}_1
    \int_{\rm bin} \mathrm{d}^3\boldsymbol{q}_2
    \int_{\rm bin} \mathrm{d}^3\boldsymbol{q}_3\,
    \delta_D^{(3)}
    \left(
    \boldsymbol{q}_1+\boldsymbol{q}_2+\boldsymbol{q}_3
    \right) .
\end{aligned}
\end{equation}

Since the 3-D Dirac delta function could be separated into a 1-D term and a 2-D term, i.e.  
\begin{equation}
    \delta_D^{(3)}
    \left(
    \boldsymbol{q}_1+\boldsymbol{q}_2+\boldsymbol{q}_3 
    \right)
    =
    \delta_D^{(1)}
    \left( q_{1\parallel}+q_{2\parallel}+q_{3\parallel}
    \right)
    \delta_D^{(2)}
    \left(
\boldsymbol{q}_{1\perp}
+\boldsymbol{q}_{2\perp}
+\boldsymbol{q}_{3\perp}
    \right),
\end{equation}
the configuration volume $V_Q$ can be separated into a parallel part and a perpendicular part,
\begin{equation}
    V_Q = V_\parallel V_\perp .
\end{equation}

For the line-of-sight part,
\begin{equation}
\begin{aligned}
    V_\parallel
    &=
 \int_{k_{1\parallel}-\Delta k_{1\parallel}/2}^{k_{1\parallel}+\Delta k_{1\parallel}/2}
\mathrm{d}q_{1\parallel}
\int_{k_{2\parallel}-\Delta k_{2\parallel}/2}^{k_{2\parallel}+\Delta k_{2\parallel}/2}
\mathrm{d}q_{2\parallel}
    \int_{-\infty}^{+\infty} \mathrm{d}q_{3\parallel}\,
    \delta_D^{(1)}
    \left(
q_{1\parallel}+q_{2\parallel}+q_{3\parallel}
    \right)            \\
    &=
\int_{k_{1\parallel}-\Delta k_{1\parallel}/2}^{k_{1\parallel}+\Delta k_{1\parallel}/2}
\mathrm{d}q_{1\parallel}    \int_{k_{2\parallel}-\Delta k_{2\parallel}/2}^{k_{2\parallel}+\Delta k_{2\parallel}/2} \mathrm{d}q_{2\parallel}
    \\
    &=\Delta k_{1\parallel}\Delta k_{2\parallel}.
\end{aligned}
\end{equation}
It means that $q_{3\parallel}$ is fixed by the triangle closure condition once $q_{1\parallel}$ and $q_{2\parallel}$ are chosen.

For the transverse part,
\begin{equation}
\begin{aligned}
    V_\perp
    &=
    \int_{\rm annuli}
\mathrm{d}^2\boldsymbol{q}_{1\perp}    \mathrm{d}^2\boldsymbol{q}_{2\perp}
\mathrm{d}^2\boldsymbol{q}_{3\perp}\,
    \delta_D^{(2)}
    \left(    \boldsymbol{q}_{1\perp}
+\boldsymbol{q}_{2\perp}    +\boldsymbol{q}_{3\perp}
    \right).
\end{aligned}
\end{equation}
Using the Fourier representation of the two-dimensional Dirac delta function,
\begin{equation}
    \delta_D^{(2)}
    \left(  \boldsymbol{q}_{1\perp}
+\boldsymbol{q}_{2\perp}    +\boldsymbol{q}_{3\perp}
    \right)
    =
    \int \frac{\mathrm{d}^2\boldsymbol{x}}{(2\pi)^2}
    \exp
    \left[
 i\,\boldsymbol{x}\cdot
    \left(    \boldsymbol{q}_{1\perp}
+\boldsymbol{q}_{2\perp}    +\boldsymbol{q}_{3\perp}
    \right)
    \right],
\end{equation}
and writing
$\mathrm{d}^2\boldsymbol{x}=x\,\mathrm{d}x\,\mathrm{d}\theta_x$ and
$\mathrm{d}^2\boldsymbol{q}_{i\perp}=q_{i\perp}\,\mathrm{d}q_{i\perp}\,\mathrm{d}\theta_i$,
we obtain
\begin{equation}
\begin{aligned}
    V_\perp
    &=
    \frac{1}{(2\pi)^2}
    \int_0^\infty x\,\mathrm{d}x
    \int_0^{2\pi}\mathrm{d}\theta_x
    \prod_{i=1}^{3}
    \left[
    \int_{k_{i\perp}-\Delta k_{i\perp}/2}^{k_{i\perp}+\Delta k_{i\perp}/2}
    q_{i\perp}\,\mathrm{d}q_{i\perp}
    \int_0^{2\pi}\mathrm{d}\theta_i\,
    \exp
    \left(
    i x q_{i\perp}\cos(\theta_i-\theta_x)
    \right)
    \right].
\end{aligned}
\end{equation}
The angular integral gives the zeroth-order Bessel function,
\begin{equation}  \int_0^{2\pi}\mathrm{d}\theta_i\,
    \exp
    \left[
    i x q_{i\perp}\cos(\theta_i-\theta_x)
    \right]
    =
    2\pi J_0(q_{i\perp}x),
\end{equation}
where
\begin{equation}
    J_0(\rho)
    =
    \frac{1}{2\pi}
    \int_0^{2\pi}
    \exp(i\rho\cos\theta)\,\mathrm{d}\theta .
\end{equation}
Therefore,
\begin{equation}
\begin{aligned}
    V_\perp
    &=
    \frac{(2\pi)^3}{(2\pi)^2}
\int_0^{2\pi}\mathrm{d}\theta_x
    \int_{k_{1\perp}-\Delta k_{1\perp}/2}^{k_{1\perp}+\Delta k_{1\perp}/2}    q_{1\perp}\,\mathrm{d}q_{1\perp}
    \int_{k_{2\perp}-\Delta k_{2\perp}/2}^{k_{2\perp}+\Delta k_{2\perp}/2}
q_{2\perp}\,\mathrm{d}q_{2\perp}              \\
    &\quad \times
    \int_{k_{3\perp}-\Delta k_{3\perp}/2}^{k_{3\perp}+\Delta k_{3\perp}/2} q_{3\perp}\,\mathrm{d}q_{3\perp}
    \int_0^\infty x\,\mathrm{d}x\,
J_0(q_{1\perp}x)J_0(q_{2\perp}x)J_0(q_{3\perp}x).
\end{aligned}
\end{equation}
The remaining radial integral is (c.f. Chapter 6-7 of \cite{GradshteynRyzhik2007})
\begin{equation}
\int_0^\infty x\,\mathrm{d}x\,
J_0(ax)J_0(bx)J_0(cx)
=
\begin{cases}
\dfrac{1}{2\pi\Delta(a,b,c)},
&
|a-b|<c<a+b, \\[1.2ex]
0,
&
0<c\leq |a-b| \ \mathrm{or}\ c\geq a+b,
\end{cases}
\label{bessel}
\end{equation}
where $\Delta(a,b,c)$ is the area of the triangle with side lengths
$a$, $b$, and $c$,
\begin{equation}
    \Delta(a,b,c)
    =
    \frac{1}{4}
    \sqrt{
    \left[(a+b)^2-c^2\right]
    \left[c^2-(a-b)^2\right]
    } .
\end{equation}
Thus the perpendicular volume is
\begin{equation}
\begin{aligned}
    V_\perp
    &=
    2\pi
    \int_{k_{1\perp}-\Delta k_{1\perp}/2}^{k_{1\perp}+\Delta k_{1\perp}/2}
    \mathrm{d}q_{1\perp}
    \int_{k_{2\perp}-\Delta k_{2\perp}/2}^{k_{2\perp}+\Delta k_{2\perp}/2}
    \mathrm{d}q_{2\perp}
    \int_{k_{3\perp}-\Delta k_{3\perp}/2}^{k_{3\perp}+\Delta k_{3\perp}/2}
    \mathrm{d}q_{3\perp}\,
    \frac{q_{1\perp}q_{2\perp}q_{3\perp}}
    {\Delta(q_{1\perp},q_{2\perp},q_{3\perp})}.
\end{aligned}
\end{equation}
In the thin-bin limit, where $\Delta k_{i\perp}\ll k_{i\perp}$ and the triangle area varies slowly across the bin, we can approximate
\begin{equation}
    q_{i\perp}\simeq k_{i\perp},
    \qquad
    \Delta(q_{1\perp},q_{2\perp},q_{3\perp})
    \simeq
    \Delta(k_{1\perp},k_{2\perp},k_{3\perp})
    \equiv \Delta_k .
\end{equation}
Then
\begin{equation}
    V_\perp
    \simeq
    2\pi
    \Delta k_{1\perp}\Delta k_{2\perp}\Delta k_{3\perp}
    \frac{k_{1\perp}k_{2\perp}k_{3\perp}}{\Delta_k}.
\end{equation}

Combining the line-of-sight and perpendicular parts gives
\begin{equation}
    V_Q
    \simeq
    2\pi\,
    \Delta k_{1\parallel}\Delta k_{2\parallel}
    \Delta k_{1\perp}\Delta k_{2\perp}\Delta k_{3\perp}
    \frac{k_{1\perp}k_{2\perp}k_{3\perp}}{\Delta_k}.
\end{equation}
If the three perpendicular $k$-bins have the same width, i.e. 
$\Delta k_{1\perp}=\Delta k_{2\perp}=\Delta k_{3\perp}\equiv \Delta k_\perp$,
this reduces to
\setlength{\fboxsep}{6pt}
\begin{equation}
\boxed{V_Q
    \simeq
    2\pi\,
    \Delta k_{1\parallel}\Delta k_{2\parallel}
    (\Delta k_\perp)^3
    \frac{k_{1\perp}k_{2\perp}k_{3\perp}}{\Delta_k}
    }.
    \label{final}
\end{equation}
Eq.\eqref{final} is the form actually used in our calculation.

\bibliography{biblio}{} 
\bibliographystyle{JHEP} 
\FloatBarrier 

\end{document}